%% file: main4.tex
\documentclass[
	aps,
	twocolumn,
	altaffilletter,
	nolongbibliography,
	numerical,
	flushbottom,
	secnumarabic,
	pra,
	superscriptaddress,
	floatfix,
	10pt
]{revtex4-2}

\usepackage{lipsum}
\usepackage{graphicx,subfigure}
\usepackage{braket}
\usepackage{amsmath, amssymb, amsthm}
\usepackage{newtxtext, newtxmath}
\usepackage{siunitx}[=v2]
\usepackage{booktabs}
\usepackage{comment}

\usepackage[breaklinks, pdftex, hyperfootnotes=true, pdfpagelabels, bookmarks, pageanchor]{hyperref}
\hypersetup{%
	colorlinks=true, linktocpage=true, pdfstartpage=1, pdfstartview=FitH, pdfborder={0 0 0},%
	breaklinks=true, pdfpagemode=UseNone, pageanchor=true, pdfpagemode=UseOutlines,%
	plainpages=false, bookmarksnumbered, bookmarksopen=true, bookmarksopenlevel=1,%
	hypertexnames=true, pdfhighlight=/O,
	urlcolor=blue, linkcolor=blue, citecolor=blue, %pagecolor=RoyalBlue,%
}

\DeclareMathOperator{\Tr}{Tr}

\usepackage[normalem]{ulem}
\newcommand{\addition}[1]{#1}
\newcommand{\replace}[2]{#2}
\newcommand{\delete}[1]{ }
\newcommand{\maybedelete}[1]{\delete{#1}}

\begin{document}
\title{Dissipative time crystals with long-range Lindbladians}
\author{Gianluca Passarelli}
\email{gianluca.passarelli@spin.cnr.it}
\affiliation{CNR-SPIN, c/o Complesso di Monte S. Angelo, via Cinthia - 80126 - Napoli, Italy}

\author{Procolo Lucignano}
\affiliation{Dipartimento di Fisica ``E.\,Pancini'', Universit\`a di Napoli Federico II, Complesso di Monte S.~Angelo, via Cinthia - 80126 - Napoli, Italy}

\author{Rosario Fazio}
\affiliation{The Abdus Salam International Center for Theoretical Physics (ICTP), Strada Costiera 11, I-34151 Trieste, Italy}
\affiliation{Dipartimento di Fisica ``E.\,Pancini'', Universit\`a di Napoli Federico II, Complesso di Monte S.~Angelo, via Cinthia - 80126 - Napoli, Italy}

\author{Angelo Russomanno}
\affiliation{Scuola Superiore Meridionale, Università di Napoli Federico II, Largo San Marcellino 10, I-80138 Napoli, Italy}

\begin{abstract}
   
    Dissipative time crystals can appear in spin systems, when the $Z_2$ symmetry of the  Hamiltonian is broken by the environment, and  the square of total spin operator $S^2$ is conserved. In this manuscript, we relax the latter condition and  show that time-translation-symmetry breaking collective oscillations  persist, in the thermodynamic limit, even in the absence of spin symmetry.
    We engineer an \textit{ad hoc} Lindbladian using power-law decaying spin operators and show that time-translation symmetry breaking  appears when the decay exponent obeys $0<\eta\leq 1$.
    This model shows a surprisingly rich phase diagram, including the time-crystal phase as well as  first-order, second-order, and continuous transitions of the fixed points. We study the phase diagram and the magnetization dynamics in the mean-field approximation. \addition{We prove that this approximation is \replace{exact}{quantitatively accurate}, when $0<\eta<1$ and the thermodynamic limit is taken, because} the system does not develop sizable quantum fluctuations, \replace{up to the third order cumulant expansion}{if the Gaussian approximation is considered}.

\end{abstract}
\maketitle

\section{Introduction}
\label{sec:intro}

Spontaneous symmetry breaking is a cornerstone of physics occurring at the most diverse energy scales, from cosmology and high-energy 
physics to condensed matter, just to mention some relevant cases. Thermal or quantum fluctuations can drive a system into a state that 
breaks, in the thermodynamic limit, some of the symmetries present in its (thermo)-dynamical potentials~\cite{Goldenfeld-book,Sachdev-book}.
Time-translational symmetry can also be spontaneously broken~\cite{wilkzek_2012}, as first conjectured by Wilczek,  leading to the existence 
of time crystals (TCs).  The mere definition of spontaneous breaking of time-translational invariance  prompted an immediate and intense
discussion~\cite{li_2012,bruno_2013,nozieres_2013,volovik_2013,syrwid_2017}.  Since then, the interest on the topic has grown enormously.
A comprehensive review of this activity can be found in Refs.~\cite{sacha_2017,khemani_2019}.
%A no-go theorem, derived in, lead
%to the conclusion that systems in thermal equilibrium cannot manifest any time-crystalline behaviour.

Time-crystal ordering   can occur only with nonlocal Hamiltonians~\cite{Kozin:PRL2019} or under nonequilibrium conditions~\cite{watanabe_2015}. A key step in this direction has been 
achieved in Refs.~\cite{else_2016,khemani_2016a} where Floquet time crystals~\cite{else_2016}  were introduced. Here a unitary system, 
subject to an external periodic driving, has observables whose expectations break the discrete time-translational 
symmetry imposed by the external drive. Floquet time crystals were intensively theoretically explored, see e.\,g. Refs.~\cite{von-keyserlingk_2016,khemani_2016,
else_2017,yao_2017,ho_2017,huang_2017, russomanno_2017}, and recently experimentally observed~\cite{zhang_2017,choi_2017}. 

\footnotetext[1]{The laser can be 
 considered classical because this phenomenon occurs also if any quantum correlation between matter and radiation can be ignored by means 
 of the adiabatic approximation.}

The time-crystal phase can also be realized in many-body open quantum systems~\cite{iemini-prl-2018:boundary-time-crystals,buca_2019,gong_2018,
tucker-newjp-2018:shattered-time, shammah-pra-2018:open-quantum-systems-with-local-and-collective-incoherent-processes,zhu_2019, lledo-pra-2019,
carollo-prl-2020:quantum-engine-time-translation-symmetry-breaking,riera-campeny-2019,seibold-pra-2020,iemini-prb-2021:btc-collective-d-level-systems,
bonaiuto-prl-2021:correlations-emitter-waveguide,piccitto-prb-2021:generalized-spin-models,hajdusek_2022,sarkar_2022,krishna_2022} where the competition 
between quantum  driving and dissipation can give rise to persisting oscillations of a collective  observable. For Markovian dynamics, the possibility of spontaneous time-translation 
symmetry breaking, in the steady state is embedded  in the properties of the Lindbladian spectrum and its scaling with the system 
size~\cite{iemini-prl-2018:boundary-time-crystals,buca_2019,shammah-pra-2018:open-quantum-systems-with-local-and-collective-incoherent-processes,
riera-campeny-2019}. In the thermodynamic limit, the real part of the Lindbladian spectrum becomes gapless, giving rise to persistent oscillations related 
to the imaginary part of the corresponding  eigenvalues. Dissipative time crystals can also be seen as cases in which only a macroscopic portion of a system (the boundary) undergoes symmetry-breaking when the remaining degrees of freedom (the bulk) acts as an effective nonequilibrium bath. We address these systems as
{\em boundary time crystals} (BTCs) as the boundary behaves effectively as a dissipative (or open) quantum system, possibly described by a Lindblad equation. 
The BTC phase has also been observed for $Z_2$-symmetric generalized $(p, q)$-spin models in the presence of collective 
dissipation~\cite{piccitto-prb-2021:generalized-spin-models}, while the TC phase does not appear when the Hamiltonian is not parity invariant~\cite{wang-pra-2021:dissipative-phase-transitions-p-spin}. More recently, the origin of BTCs has been attributed to the Lindbladian 
steady-state being parity-time symmetric~\cite{nakanishi-2022:btc-from-pt-symmetry}.
Most notably, in the case of open-system dynamics it is possible to realize continuous time crystals, whose first experimental 
implementation has been reported in Ref.~\cite{kesler_2022}. 

Despite the vibrant scientific activity surrounding this topic, many aspects of time crystals in open systems remain yet to be fully understood. 
The spontaneous generation of a collective periodic oscillation in classical dissipative systems has a long history. Examples in this sense are, for instance, the synchronization in the Kuramoto model~\cite{Kuku,Kuku1}, the laser~\cite{RevModPhys.47.67,Note1}, the salt oscillator~\cite{salty} and, in some sense, also the Belusov-Zhabotinski reaction~\cite{Kapral:book}. 
Moreover, already in the 90's, Refs.~\cite{PhysRevLett.70.3607,PhysRevA.41.1932} provided an extensive analysis of when the generation 
of subharmonics can and cannot occur in the case of many-body classical dissipative driven systems (see also Ref.~\cite{machado}). The great interest in time crystals has brought new examples of this sort~\cite{Yao_2020,Gambetta_2019,
pizzi2020bistability}, opening also the possibility for a deeper understanding of possible connection with these phenomena.

A very useful framework to understand  time crystals in open systems is to link their existence to the emergence of decoherence-free 
subspaces~\cite{zanardi_1997,lidar_1998} in the thermodynamic limit. It is thus natural to expect that symmetries in the coupling to the external bath should play a major role. Indeed all spin models supporting dissipative TCs so far implicitly assume a collective (infinite-range) coupling to the external environment. 

In this work we make one step further, and introduce a new class of long-range dissipators that spatially decay as a power law and allow nevertheless a BTC phase. 
%In our opinion, this is relevant for several different reasons. First of all, 
The resulting phase diagram is very rich as a function of the range of the  dissipation and the coupling parameters, with both first-order, second-order, and continuous transitions of the fixed points, as well as a coexisting region.

BTC phases are believed to appear in  spin models in the presence of two fundamental ingredients, that are (i) a $Z_2$ symmetry of the Hamiltonian 
that is explicitly broken by the environment~\cite{piccitto-prb-2021:generalized-spin-models}, and (ii) ``strong'' rotational 
symmetry~\cite{booker-njp-2020:btc-spectral-properties-and-finite-size-effects}, %\gian{Leggere con attenzione}
which allows to decouple the eigenspaces of the total angular momentum operator. Here, ``strong'' means that both the Hamiltonian and the 
Lindblad operators are functions of collective spin operators~\cite{piccitto-prb-2021:generalized-spin-models}, expressed as {\em uniform} sums 
of onsite spins. As a result, the square of the total angular momentum ${S}^2 = {S}_x^2 + {S}_y^2 + {S}_z^2$ is conserved at the operator level 
for any finite size.  

In this paper we inquire the role of condition (ii). In particular we break it and study if the boundary time crystal is still present. We focus on 
the Lindblad operators, and substitute the collective spins with power-law decaying operators. 
%of Ref.~\cite{iemini-prl-2018:boundary-time-crystals}   
Our finding is that condition (ii) is not strictly necessary in order to have a boundary time crystal.
In analogy with the Hamiltonian case~\cite{Khasseh:prl2019,Pizzi:natcomm2021,russomanno_2017} also here the time-crystal phase is  supported up to a critical value of the exponent governing the power-law decay.
This is the first example of a phase transition engineered by tuning the range of dissipation. 

The paper is organized as follows. In Sec.~\ref{sec:model}, we introduce our model Lindbladian with power-law decaying Lindblad operators and discuss
its general properties. In Sec.~\ref{sec:mean-field}, we resort to mean-field theory to derive the equations of motion of the magnetization components 
and we use them to study both the dynamics and the phase diagram of our model. In Sec.~\ref{sec:third-cumulant} we discuss the dynamics 
of the magnetization at the third order in the cumulant expansion and show that correlations provide only a small correction to the mean-field dynamics 
when the power-law exponent is below a threshold, in the thermodynamic limit. We finally summarize and comment on our results in Sec.~\ref{sec:conclusions}. Some technical aspects of our analysis 
are discussed in a number of dedicated Appendices.

%.......................................................................................................................................%
\section{Model and phase diagram}
\label{sec:model}

\addition{The simplest model providing a boundary time crystal phase is given by the following Lindbladian, first proposed and studied in Ref.~\cite{iemini-prl-2018:boundary-time-crystals}:}
\begin{equation}\label{eq:btc}
    \dot{\rho} = -i \left[ 2 J \, S_x, \rho \right] + \frac{\gamma}{N} \left(S_+ \rho S_- - \frac{1}{2} \left\lbrace S_- S_+, \rho \right\rbrace  \right),
\end{equation}
\addition{where}
\begin{equation}\label{scoll:eqn}
    S_\alpha = \frac{1}{2}\sum_{i=1}^{N} \sigma_i^\alpha, \quad \alpha \in \left\lbrace x, y, z\right\rbrace
\end{equation}
\addition{are collective spin operators with algebra $ [S_\alpha, S_\beta] = i \, \epsilon_{\alpha\beta\gamma} S_\gamma $, $ S_\pm = S_x \pm i S_y $, and $ S $ is the total spin. The $\sigma_i^\alpha$ are Pauli matrices and $\sigma_i^\pm = (\sigma_i^x \pm i \sigma_i^y)/2$. The number of particles is $N$, and the components of the magnetization are defined as $ m_\alpha = 2\langle S_\alpha \rangle / N $, where the expectation is taken over the density operator $\rho$.}

\addition{The Hamiltonian part describes noninteracting (free) spins in a uniform magnetic field oriented in the $x$ direction. The resulting Hamiltonian, as seen in the first term at the r.\,h.\,s. of Eq.~\eqref{eq:btc}, is $H = 2J \, S_x $. This Hamiltonian is time-independent: the bare system is invariant by continuous time translations. The second term in the r.\,h.\,s. of Eq.~\eqref{eq:btc} is the environment, acting by orienting the spins in the $z$ direction, towards the state with a positive magnetization. If $S_+$ and $S_-$ are exchanged, the model still features a BTC phase.}

\addition{The Hamiltonian and the jump operators commute with $S^2 = S_x^2 + S_y^2 + S_z^2$, and the conditions (i) and (ii), mentioned in Sec.~\ref{sec:intro}, are satisfied, indeed this model possesses a TC phase. This phase exists in the thermodynamic limit and can be analyzed using mean-field theory, since, for large $N$, correlations between collective variables vanish as $1/N$, $ [S_\alpha, S_\beta]/N^2 = O(1/N) $, and the magnetization behaves like a classical variable. The dissipative phase diagram of the model features a critical point $\chi = \gamma/4J = 1$ separating the boundary time crystal phase for weak dissipation ($\chi < 1$) from an ordered magnetic phase where the spin state is magnetized ($\chi > 1$) and the $Z_2$ symmetry is manifestly broken.}

Motivated by Refs.~\cite{seetharam_2022a,seetharam_2022b,marino:long-range-universality} we consider a generalization of \replace{the model considered in Ref.~\cite{iemini-prl-2018:boundary-time-crystals} 
(see Appendix~A}{this model}. We consider a system of $N$ spin-$1/2$ particles described by Pauli matrices $\sigma_i^\alpha $ with 
$ \alpha \in \left\lbrace x, y, z\right\rbrace $.
The dynamics is governed by a Lindblad equation
	\begin{equation}\label{eq:lindbladian-interpolated}
        \dot{\rho} = -i \left[ H, \rho \right] + \gamma \!\sum_{i=1}^N \left(L_i(\eta) \rho L_i^\dagger(\eta) - \frac{1}{2} \left\lbrace L_i^\dagger(\eta) L_i(\eta), \rho \right\rbrace  \right).
    \end{equation}
 We take the same Hamiltonian as in Ref.~\cite{iemini-prl-2018:boundary-time-crystals}, $H=2J \,{S}_x$ and the following Lindblad operators,
\begin{equation}\label{eq:power-law-lindblad}
		L_i(\eta) = \sum_{j=1}^N f_{ij}(\eta) \, \sigma_j^+, \quad f_{ij}(\eta) = \frac{K^{(N)}(\eta)}{{D(\lvert i- j\rvert)}^\eta},
	\end{equation}
	where \delete{$\sigma_i^\pm = (\sigma_i^x \pm i \sigma_i^y)/2$,}$\eta \in [0, \infty) $ is the power-law exponent, and $D(r)$ is a distance function between lattice sites defined as $ D(r) = \min(r, N-r) + 1 $ in order to provide periodic boundary conditions~\cite{gor,parisi:sr-lr-ising,piccitto:long-range,halimeh2,gong:locality-power-law,passarelli:cd-closed,ren:entanglement-long-range}. The Kac normalization factor $K^{(N)}(\eta)$ ensures that $\sum_{j=1}^N f_{ij}(\eta) = 1$~\cite{kac-jmathphys-1963:van-der-waals} and, for this choice of $D(r)$, it can be computed analytically for any $\eta$ (see Appendix~\ref{app:kac} for details). Notice that $f_{ij}(\eta)=f_{\lvert i-j\rvert}(\eta)$.

A dissipative term in the Lindbladian with power-law decaying correlation can be realized, for example, in cold atoms inside a cavity. As discussed in Ref.~\cite{seetharam_2022a}, a key element in this proposal is the presence of three-level atoms, trapped inside an optical cavity. In the presence of  a magnetic field gradient, and a Raman beam,  several sidebands of tunable frequency and amplitude appear, that make pairs of atoms at long distance apart interact. In these conditions, one can show that the dynamics of the atoms is described by a Lindblad equation, with long-range Lindbladians of the form of our Eq. (4). A detailed analysis of the protocol leading to the power-law dissipator is described in Appendix~A of Ref.~\cite{seetharam_2022a}. 

The operators in Eq.~\eqref{eq:power-law-lindblad} are sums of local spins, and these sums are not uniform, but have a coefficient decaying with distance as a power law, with exponent 
$\eta$. In this way, for any finite size and any $\eta>0$, ${S}^2$ is not conserved.%, although the model is still \replace{symmetric under all the possible site permutations.
%We emphasize that the Lindbladian in Eq.~\eqref{eq:lindbladian-interpolated} is symmetric with respect to any permutation of spin lattice indices. 
%In fact, not only is the Hamiltonian $H$ permutation invariant, but also the Lindblad operators preserve the permutation invariance of the model for any given $\eta$, despite breaking the conservation of ${S}^2$. The starting state has to be symmetric under all spin permutations}{translationally invariant}. %This symmetry is crucial for the forthcoming discussion.

%We label the spins using indices in $\set{1,2,\dots,N}$, and use them to evaluate the distance function $D(r)$, similarly to Hamiltonians with power-law interactions (see Refs.~\cite{parisi:sr-lr-ising,piccitto:long-range,halimeh2,gong:locality-power-law,passarelli:cd-closed,ren:entanglement-long-range} for some recent studies).
	
	Let us first focus on two limiting cases.
    For $\eta = 0$, the so-called infinite-range case, we have that
    \begin{equation}
        f_{ij}(0) = \frac{1}{N}, \qquad L_i(0) = \frac{S_+}{N} \qquad \forall i, j:
    \end{equation}
    here, the Lindbladian of Eq.~\eqref{eq:lindbladian-interpolated} coincides with the collective one of Eq.~\eqref{eq:btc}.
    In the opposite, zero-range limit we have that
    \begin{equation*}
        f_{ij}(\eta\to\infty) = \delta_{ij}, \qquad L_i(\eta\to\infty) = \sigma_i^+:
    \end{equation*}
    the corresponding Lindbladian acts independently on each spin, thus the zero-range Lindbladian cannot generate correlations during the open-system evolution. In some cases, collective and independent baths give rise only to quantitative changes in the physics~\cite{Passarelli:pra2020,Passarelli:pra2022}, but there are also many examples where these two kinds of dissipation give rise to qualitative differences (e.\,g., superradiance in the Dicke model~\cite{shammah-pra-2018:open-quantum-systems-with-local-and-collective-incoherent-processes}). The model discussed in this paper falls in the latter category, because the two limiting cases are qualitatively different. %While, in most cases, collective and independent baths give rise only to quantitative changes in the physics~\cite{Passarelli:pra2020,Passarelli:pra2022}, in our model the two limiting cases are qualitatively different. 
    
    The Lindbladian of Eq.~\eqref{eq:power-law-lindblad} interpolates between the infinite-range one of Eq.~\eqref{eq:btc} and a local Lindbladian where each spin is independently coupled to its own environment via $\sigma_i^+$. For every choice of $\eta$, the Lindblad operators act by orienting the spins in the $z$-direction, thus breaking the $Z_2$ symmetry. For the two limiting cases, $\eta = 0$ and $\eta = \infty$, correlations are negligible: Here, the mean-field approximation is (for different reasons) \replace{exact}{quantitatively accurate}. 
    
    For intermediate values of $\eta$ less is known. In the following, we are going to show that quantum correlations are negligible also for $0<\eta\leq 1$, in the thermodynamic limit. In this regime the dynamics of the magnetization components is well described by the mean-field theory.
    
\begin{figure}[t]
        \centering
        \includegraphics[width=\columnwidth]{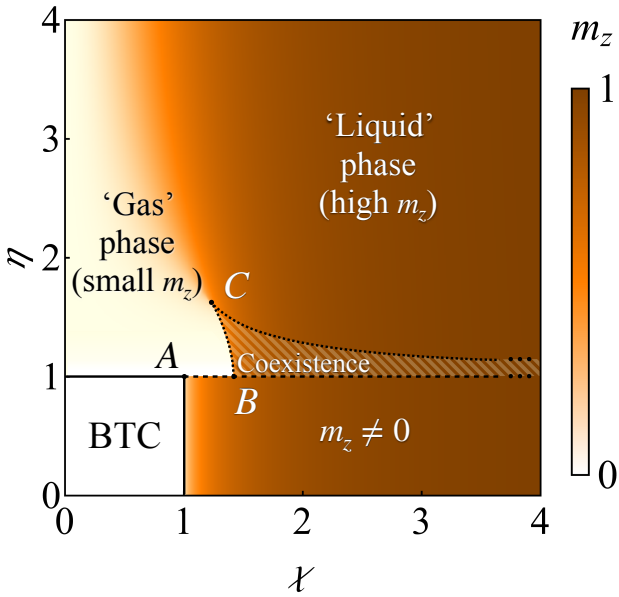}
        \caption{Sketch of the phase diagram of the model studied in this paper. The boundary time crystal phase is observed for $0\le\eta\le1$ and $\chi\le1$. For $\eta \le 1$, the phase transition to the magnetized phase is independent of $\eta$. For $\eta > 1$, the BTC phase no longer exists. For $1 < \eta \lesssim 1.625$, the system undergoes a first-order phase transition as a function of $\chi$, separating a phase where the steady-state magnetization is close to zero without featuring persistent oscillations and a coexistence phase where there are two stable fixed points with different magnetization. For $\eta>1.626$ the phase transition disappears and the system qualitatively resembles its zero-range limit ($\eta = \infty$). The three points marked in the figure are $A = (1,1)$, $B=(\sqrt{2}, 1)$, and $C = (1.225,1.625)$.}
        \label{fig:sketch}
    \end{figure}

   Before moving to the next section, we would like to summarize the main results of this paper. The  mean-field phase diagram of our model is very rich. We can plot it, taking as parameters the decay exponent $\eta$ and the normalized strength of the dissipator $\chi = \gamma / 4J$. We show the phase diagram in Fig.~\ref{fig:sketch}. The two regimes $0 < \eta \le 1$ and $\eta > 1$ will be loosely denoted long-range and short-range, respectively. 
   
   In the long-range regime, we see a transition from a time-crystal phase to a {phase with an asymptotic nonvanishing $z$-magnetization}. Notice that the transition point in $\chi$ does not depend on $\eta$ in this interval, and we find here collective persistent oscillations in the thermodynamic limit for $\chi < 1$. Quite remarkably, in this time-crystal regime, the dynamics conserves the expectation of ${S}^2$ in the 
thermodynamic limit but not for finite system sizes. 

Our results set up an interesting parallel with Hamiltonians with long-range interactions. 
In that case, in dimension $d=1$, the unitary dynamics is equivalent to the case $\eta=0$ whenever $0<\eta\le1$
~\cite{pappalardi-prb-2018:scrambling-and-entanglement-spreading,halimeh1,halimeh3,zunkovic-prl-2018:dynamical-quantum-phase-transitions,cevolani-newjphys-2016:spreading-of-correlations}, 
and we find the same here for power-law dissipators.

   For $ \eta > 1$, the magnetization always relaxes to an asymptotic value, and there are no time-crystal oscillations. The line $\eta = 1$ separating these two regimes is a phase-transition line. The nature of the phase transition depends on the value of $\chi$. For $\chi < 1$, there is a continuous phase transition between the vanishing-magnetization fixed point of the BTC phase and a phase with small but nonzero magnetization. At the critical line, the magnetization is infinitely differentiable, but nonanalytic.
    For $\chi > 1$, the line at $\eta=1$ marks a first-order discontinuous phase transition.
    
    For $\eta>1$, we observe a region where there are two coexisting stable fixed points: one corresponding to a small steady-state magnetization (similar to a low-density, ‘gas’ phase) and one to a large magnetization (similar to a high-density, ‘liquid’ phase). The coexistence region ends at the critical point $C = (\chi_C, \eta_C) = (1.225, 1.625)$. The line $BC$ is a first-order critical line; above the critical point $C$, there are no phase transitions and the steady state is always unique, smoothly interpolating the gas and liquid phases as a function of $\chi$, as occurs in the limit $\eta \to \infty$. A similar first-order transition line ending with a critical point, separating a low-density gas phase from a high-density liquid phase, can be seen in the equilibrium phase diagram of water~\cite{Goldenfeld-book}. More in general, this is an example of imperfect pitchfork bifurcation~\cite{Strogatz}, as we better clarify in Sec.~\ref{sec:mean-field}.
   
    %In addition, for $1<\eta\lesssim 1.625$ there is a first-order phase transition as a function of $\chi$, separating a phase with a steady-state magnetization that is close to zero and a phase with large asymptotic $z$-magnetization. For larger values of $\eta$ there are no phase transitions and the asymptotic magnetization and its derivatives vary continuously as a function of $\chi$, qualitatively resembling the same behavior found in the zero-range limit $\eta = \infty$.
    
    In the transient regime before these asymptotic values set in, interesting effects occur for $1 < \eta\lesssim 1.625$. Here, for small values of $\chi$, the relaxation to the unique stable steady state occurs through decaying oscillations towards an asymptotic value that is close to zero. For high enough values of $\chi$, there is again a single stable steady state and the oscillations are quickly damped. Between these two regimes, there is the coexistence regime, and each of the two stable fixed points has its own basin of attraction. Initializing near the boundary between the two basins of attraction, there is a transient approach to the large-magnetization fixed point, and then damped oscillations with convergence to the small-magnetization fixed point.
    
    In the next section, we discuss the phase diagram in more detail. 
    
%..............................................................................................................................%
\section{Mean-field analysis}
\label{sec:mean-field}

In this section we discuss the results obtained with the mean-field approximation, summarized in the phase diagram of Fig.~\ref{fig:sketch}, where the parameters are the exponent $\eta$ and the normalized dissipation strength $\chi $. %For $\eta \le 1$ (below the dashed line), the equations of motion for the magnetization components $m_\alpha$ are the same as those for $\eta = 0$. The BTC phase is observed for $\chi \le 1$. For $\eta > 1$, the BTC phase is destroyed, but we can observe two different behaviors depending on $\chi$. In particular, for small values of $\chi$, the magnetization $m_z(t)$ oscillates but the amplitude of these oscillations decays over time. There is also a region where these oscillations start to appear after a time delay. When the dissipation rate becomes larger, the system enters the $Z_2$ broken phase. %It is remarkable that the time crystal phase can survive even though the Lindblad operators are not collective spin operators and spatially decay with an exponent $0 < \eta \le 1$.
    
\begin{figure}[t]
        \centering
        \includegraphics[width=0.9\columnwidth]{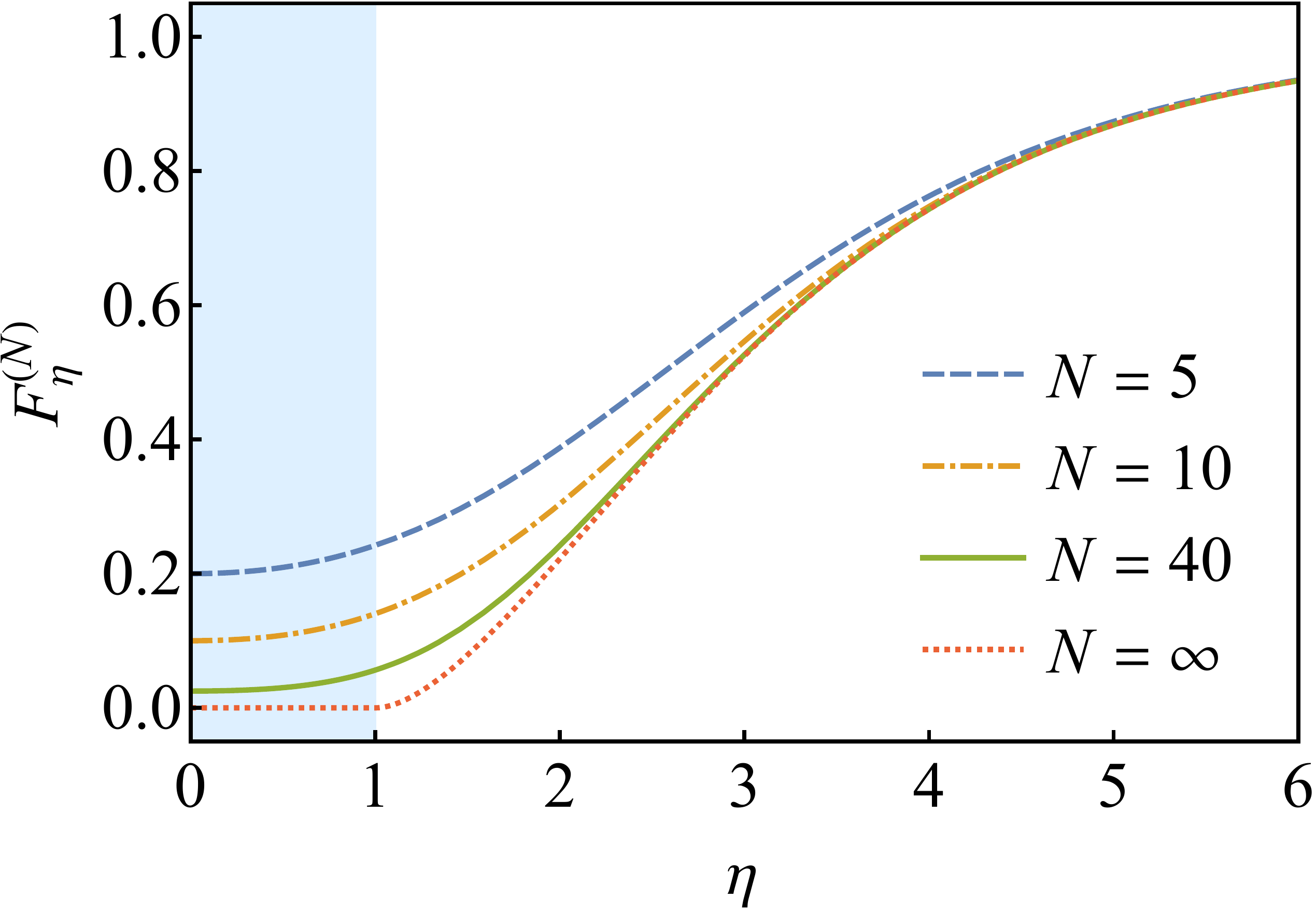}
        \caption{$F_\eta^{(N)}$ as a function of $\eta$ for several values of $N$. For $N = \infty$, $F_\eta = 0$ for $0\le\eta\le1$. This is the long-range regime (highlighted in light blue).}
        \label{fig:coeff-f1}
    \end{figure}

    The mean-field equations of motion for the magnetization components can be derived analytically by only exploiting the model's \replace{permutation}{translational} invariance, see Appendix~\ref{app:magnetization-dynamics}. In the thermodynamic limit, they read
    \begin{equation}\label{eq:power-law-lindblad-magnetization-mf}
		\begin{cases}
			\displaystyle\dot{m}_x = -\frac{\gamma}{2} m_x F_\eta -\frac{\gamma}{2} m_x m_z (1-F_\eta);\\[1.5ex]
			\displaystyle\dot{m}_y = 2J m_z -\frac{\gamma}{2} m_y F_\eta -\frac{\gamma}{2} m_y m_z (1-F_\eta);\\[1.5ex]
			\displaystyle\dot{m}_z = -2J m_y +\gamma(1-m_z) F_\eta + \frac{\gamma}{2}(m_x^2 + m_y^2) (1-F_\eta)\,,
		\end{cases}
	\end{equation}
    where, considering Eq.~\eqref{scoll:eqn}, we have defined the magnetization components as the expectations of the total spin components
\begin{equation} \label{mm:eqn}
  m_\alpha(t)=2\lim_{N\to\infty}\frac{\Tr[\rho(t) {S}_\alpha]}{N}\quad  ( \alpha \in \left\lbrace x, y, z\right\rbrace )
\end{equation}
and omitted time-dependences for shortness. The coefficient $ F_\eta $ is given by
    \begin{equation}
        F_\eta = \lim\limits_{N\to\infty} \frac{1}{N}\sum_{i,j=1}^{N} f_{ij}^2(\eta) = 
        \begin{cases}
			\displaystyle\frac{2\zeta(2\eta)-1}{{[2\zeta(\eta) - 1]}^2} & \text{$\eta > 1$,}\\
			0	& \text{$0\le\eta\le1$,}
		\end{cases}
    \end{equation}
    where $\zeta(z)$ is Riemann's Zeta function. We plot $F_\eta$ versus $\eta$ in Fig.~\ref{fig:coeff-f1}. We also plot the functions $F^{(N)}(\eta)$ for finite values of $N$. %We have different regimes corresponding to the different regions of the phase diagram in Fig.~\ref{fig:sketch}.
    
%\subsection{Absence of chaos}
%\label{sec:mf-chaos}
    
    The highlighted light-blue range $0 \le \eta \le 1$ in Fig.~\ref{fig:coeff-f1} marks the parameter region where, in the thermodynamic limit, $F_\eta = 0$. This is the so-called long-range regime: here, the power-law Lindblad operators do not affect the thermodynamic limit behavior of the coefficients of local and collective pump. In this case, since $ F_\eta = 0 $, the equations of motion [Eqs.~\eqref{eq:power-law-lindblad-magnetization-mf}] are exactly the same as the ones derived in Ref.~\cite{iemini-prl-2018:boundary-time-crystals} for the Lindbladian of Eq.~\eqref{eq:btc}. Thus, the boundary time crystal phase is not affected by the power-law decay of the jump operators in the long-range regime and exists for all values of $0\le \eta\le 1$, as one can see in Fig.~\ref{fig:sketch}. Here, the two quantities
    \begin{equation}\label{eq:conserved}
        \mathcal{N} = m_x^2 + m_y^2 + m_z^2,\qquad
        \mathcal{M} = \frac{m_x}{m_y - 1/\chi}
    \end{equation}
    are constants of motion, even if $[L_i(\eta), S^2] \ne 0$.
    
    In the next two subsections we are going to see how the properties of the system change by moving away from this range of parameters. In Sec.~\ref{fixed:sec} we consider the situation from the point of view of the fixed points of the dynamics, and in Sec.~\ref{dyno:sec} from the point of view of the dynamics.
    
%------------------------------------------------------------------------------------------------------------------------------------%
\subsection{Fixed points}    \label{fixed:sec}
 In order to find the fixed points, we impose $\dot{m}_x=\dot{m}_y=\dot{m}_z=0$ in Eq.~\eqref{eq:power-law-lindblad-magnetization-mf}. Let us start by considering the situation in the long-range regime $\eta<1$. In this case, the fixed-point solution is
	\begin{equation}\label{compo:eqn}
		\begin{array}{ccc@{\hskip 3mm}c}
			\displaystyle m_x = 0;  &\displaystyle m_y = \frac{1}{\chi}; & \displaystyle m_z = \frac{\sqrt{\chi^2-1}}{\chi} & \text{if $\chi \ge 1$},\\[1ex]
			\displaystyle m_x = \sqrt{1-\chi^2}; &\displaystyle m_y = \chi; & \displaystyle m_z = 0 & \text{if $\chi < 1$}.
		\end{array}
		\end{equation}
	At $\chi = 1$, there is a second-order phase transition between the BTC phase ($\chi < 1$) and the ordered ferromagnetic phase ($\chi > 1$), because one can see a discontinuity of $\partial_\chi m_z$ at $\chi=1$. We notice also that at $\chi=1$ the nature of the fixed point changes, from an elliptic fixed point around which the trajectories orbit, to a stable attractive fixed point onto which the trajectories fall asymptotically.

    \begin{figure*}[t]
        \centering
        \includegraphics[width=\textwidth]{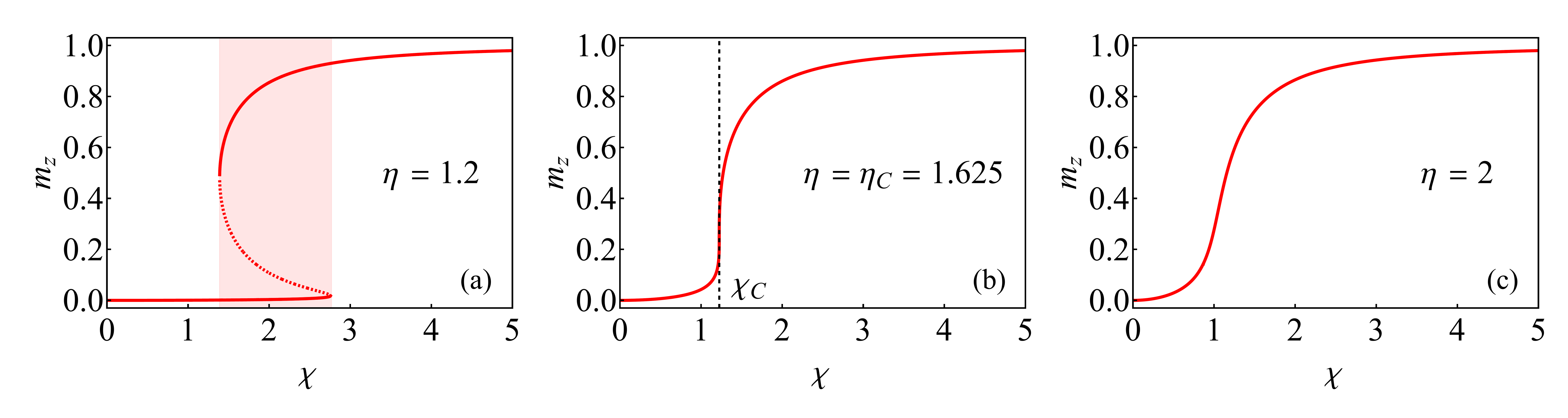}
        \caption{Steady-state magnetization as a function of $\chi$. (a): $\eta = 1.2$; (b): $\eta = \eta_C = 1.625$; (c): $\eta = 2$. For $\eta < \eta_C$, we observe the coexistence of two stable solutions in the shaded area (solid red lines) and a single unstable one (dotted red line). For $\eta = \eta_C$, the derivative $\partial_\chi m_z$ is infinite at $\chi = \chi_C = 1.225$.}
        \label{fig:hysteresis}
    \end{figure*}

    Let us now move to another simple limit, $\eta\to\infty$ (zero-range), which has been thoroughly analyzed in Ref.~\cite{maghrebi-pra-2021:driven-dissipative-ising-model}. In this limit, the Hamiltonian and the Lindblad operators act independently on single spins, thus correlations never build up in the system and the equations of motion for $m_\alpha$ can be written down exactly without even resorting to the mean-field approximation. The resulting equations for the expectation values are given by Eq.~\eqref{eq:power-law-lindblad-magnetization-mf} with $F_\eta=1$ and the steady-state solution is 
    \begin{equation}\label{eq:magnetization-local}
		m_x = 0; \quad m_y = \frac{2\chi}{2\chi^2+1}; \quad m_z = \frac{2\chi^2}{2\chi^2+1}\,.
	\end{equation}

	Here there are no critical points nor conserved quantities. We find that a qualitatively similar picture characterized by the absence of a phase transition holds whenever $\eta \gtrsim 1.625 $: in this short-range regime, the dynamics are mostly controlled by local pump processes.

	When one crosses the $\eta=1$ line from $\eta<1$ to $\eta>1$, the fixed point changes in a nonanalytic way and features therefore a phase transition. Crossing this line, there is a sudden change in the equations of motion for $m_\alpha$, due to the fact that for $\eta>1$ the two quantities in Eq.~\eqref{eq:conserved} are no longer conserved and the BTC phase is destroyed. For $\eta>1$ the fixed-point solutions to Eqs.~\eqref{eq:power-law-lindblad-magnetization-mf} is found from the following system of equations:
	\begin{equation}\label{eq:fixed-point-eta-greater-than-one}
	    \begin{cases}
	        m_x = 0\\
	        m_y = \displaystyle\frac{m_z}{\chi [F_\eta + m_z (1-F_\eta)]}\\
	        \begin{aligned}
	            &m_z^3 - (1-2\lambda) m_z^2 - \lambda \left[2-\lambda\left(1 + \frac{1}{2\chi^2 F_\eta^2}\right)\right] m_z \\ &\quad{}- \lambda^2 = \prod_{k=1}^3 (m_z - m_z^{(k)}) = 0,
	        \end{aligned}
	    \end{cases}
	\end{equation}
	where $\lambda = F_\eta / (1-F_\eta)$. Since, in this range of $\eta$, $\lambda \ne 0$, we find that $m_z^{(k)} = 0$ is never a solution to these equations.

    A special regime occurs when $1 < \eta < 1.625$. In this regime Eqs.~\eqref{eq:fixed-point-eta-greater-than-one} can either have one or three real solutions depending on the value of $\chi$. Small values of $\chi$ result in a unique gas-like steady state, while large values of $\chi$ lead to a unique liquid-like steady state. In between the gas-like and liquid-like phases, there is a coexistence region, shown in Fig.~\ref{fig:sketch} with a dashed filling pattern, where there are three real solutions to Eqs.~\eqref{eq:fixed-point-eta-greater-than-one}. The analysis of the stability of these fixed points shows that one of them is always unstable and repulsive, whereas the remaining two are stable and attractive. One of these solutions, $m_z^{(1)}$, corresponds to a steady state with large total magnetization $\mathcal{N} \approx 1$, while the other one, $m_z^{(2)}$, has a small total magnetization $\mathcal{N} \approx 0$. 
    We show an example of the fixed-point $m_z$ versus $\chi$ for $\eta = 1.2$ in Fig.~\ref{fig:hysteresis}(a). The shaded area shows the coexistence of two stable fixed points, denoted by the solid lines. The dotted line represents the third, unstable fixed point. This situation corresponds to a first-order phase transition, because $m_z$ changes discontinuously as a function of $\chi$.

    This coexistence situation disappears for $\eta > 1.625$, where the equation for the fixed-point $m_z$ always has two complex roots and a real one, the latter corresponding to a stable fixed point. The magnetization curve is therefore continuous and regular without phase transition and coexistence regime [see Fig.~\ref{fig:hysteresis}(c) for $\eta = 2$]. Exactly at $\eta = \eta_C = 1.625$, the magnetization curve is continuous and regular but at $\chi=\chi_C = 1.225$, where it has a vertical-tangent flex, as shown in Fig.~\ref{fig:hysteresis}(b). So, there is a discontinuity in $\partial_\chi m_z$, corresponding to the critical point marked as $C$ in Fig.~\ref{fig:sketch}. So, decreasing $\eta$, one has first one real stable and two complex solutions of Eq.~\eqref{eq:fixed-point-eta-greater-than-one} for all $\chi$, then a value of $\chi$ with three coinciding real solutions (the critical point), and then an interval of $\chi$ where there are two real stable solution and a real unstable one (coexistence regime). This phenomenon is standard in nonlinear dynamics and is known as imperfect pitchfork bifurcation (or cusp catastrophe)~\cite{arovas,Strogatz}. 
 
    \begin{figure}[t]
        \centering
        \includegraphics[width=0.9\columnwidth]{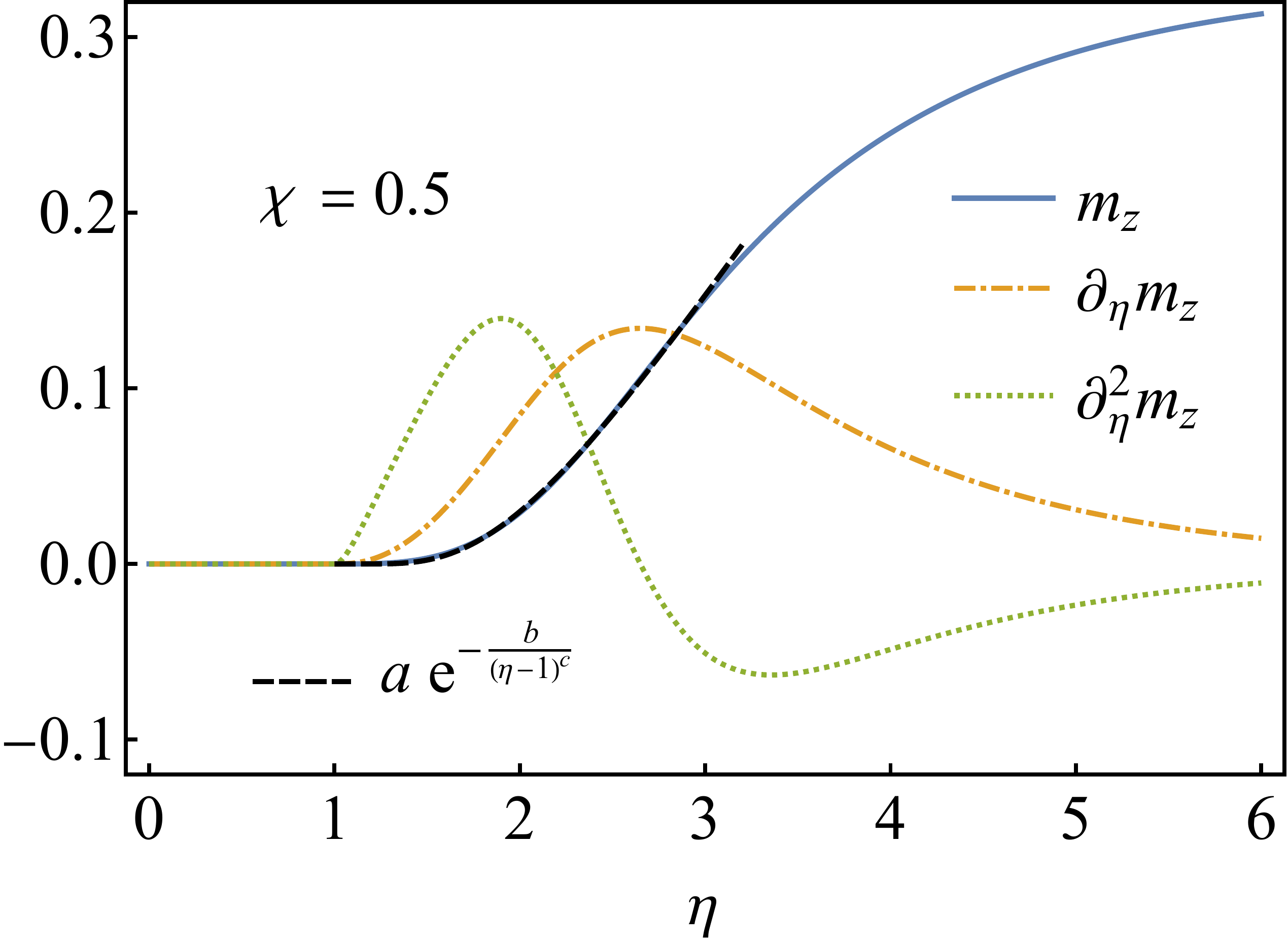}
        \caption{Magnetization $m_z$ and its derivatives as a function of $\eta$ for $\chi = 0.5$. The curve for $m_z$ is well described by the function in Eq.~\eqref{fit:eqn}, which is nonanalytic in $\eta=1$ (best-fit parameters: $a=2.5$, $b=4.4$, $c=0.66$).}
        \label{fig:no-phase-transition-btc}
    \end{figure}
 
 \begin{figure*}[t]
        \centering
        \includegraphics[width=\textwidth]{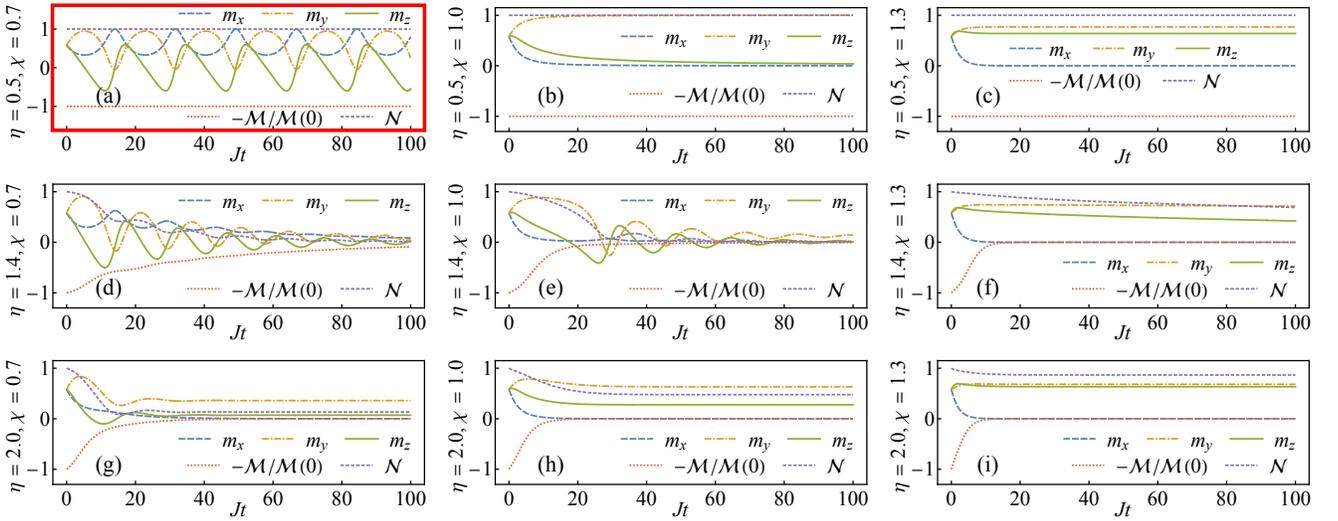}
        \caption{Dynamics of the magnetization components, $m_\alpha$, and of the two functions $\mathcal{M}$ and $\mathcal{N}$ described in the main text, for several choices of the exponent $\eta$ and the dissipation rate $\chi$. Top row: $\eta = 0.5$; center row: $\eta = 1.4$; bottom row: $\eta = 2.0$. Left column: $\chi = 0.7$; center column: $\chi = 1.0$; right column: $\chi = 1.3$. In panel (a), highlighted with a bold red frame, we observe a BTC phase. For $\eta = 0.5$, $\mathcal{M} $ and $\mathcal{N}$ are conserved quantities. Technical details:  dynamics numerically performed with implicit backward differentiation formulas~\cite{hindmarsh2005sundials}, initial condition $m_\alpha(0) = 1/\sqrt{3}$ $\forall\alpha$.}
        \label{fig:dynamics}
    \end{figure*}   
    
     Let us now comment on how the fixed points change when the $\eta=1$ line is crossed. For $\chi>1$ they change in a discontinuous way, featuring a first-order transition. For $\chi < 1$ the transition is between the BTC phase ($\eta<1$) and the gas-like phase ($\eta>1$) and is a continuous $\infty$-order transition for the fixed point. Indeed, crossing the $\eta=1$ line, the fixed-point $m_z$ is infinitely differentiable with continuous derivatives but is not analytic. This can be seen in Fig.~\ref{fig:no-phase-transition-btc}, where we plot the fixed-point magnetization and its first and second derivative as a function of $\eta$ for $\chi = 0.5$. We see that $m_z$ is well fitted around $\eta = 1$ by a function of the form
    \begin{equation} \label{fit:eqn}
        m_z(\eta) = a e^{-\frac{b}{{(\eta-1)}^c}},
    \end{equation}
    where $a$, $b$ and $c$ are fitting parameters.

In the next subsection we discuss the dynamics around these fixed points, leading to persistent oscillations in the BTC phase, and to relaxation to an attractive fixed point otherwise, with interesting observations when two attractive fixed points coexist.

%------------------------------------------------------------------------------------------------------------------------------------%

    \subsection{Dynamics} \label{dyno:sec}
    
    \textit{BTC phase.} Let us start by considering the BTC phase. Here the time crystal supported by the long-range power-law Lindblad operators with $\eta \le 1$ is the same as the one studied in Ref.~\cite{iemini-prl-2018:boundary-time-crystals}, i.\,e., a mean-field semiclassical time crystal. We can see some time traces for $\eta=0.5$ in Figs.~\ref{fig:dynamics}(a--c). Fig.~\ref{fig:dynamics}(a) corresponds to the time-crystal phase, and one can see the persisting oscillations of the magnetization components. Figs.~\ref{fig:dynamics}(b,\,c) correspond to the symmetry-broken phase for $\chi>1$, and one can see the magnetization components reaching the asymptotic value given by Eq.~\eqref{compo:eqn}. In all these cases, $\mathcal{M}$ and $\mathcal{N}$ are conserved.
    \begin{figure}
        \centering
        \includegraphics[width=0.95\columnwidth]{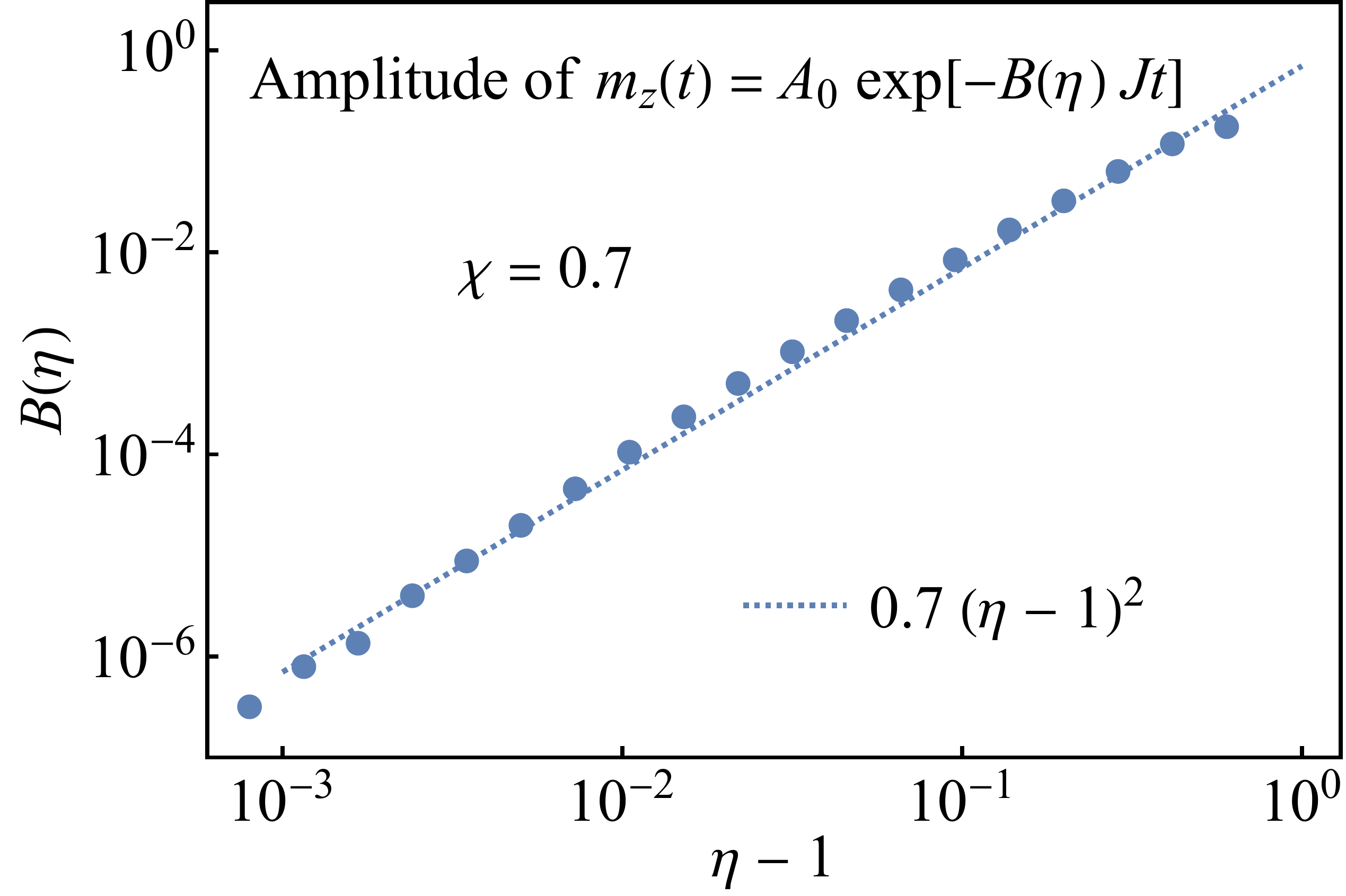}
        \caption{Decay rate of the amplitude of $m_z(t)$ for $\eta = 1.1$ and $\chi = 0.7$.}
        \label{fig:metastable}
    \end{figure}
    
    \textit{Short-range regime.} Some examples of dynamics in the short-range regime $\eta > 1$ are shown in Figs.~\ref{fig:dynamics}(d--i). In Figs.~\ref{fig:dynamics}(d,\,e) we see oscillations decaying to a value close to zero, in Fig.~\ref{fig:dynamics}(f) the magnetization relaxes to an asymptotic value in the liquid phase. Both $\mathcal{M}$ and $\mathcal{N}$ [see Eq.~\eqref{eq:conserved}] are not conserved and decay.
    For $\eta = 2$, the oscillations of the magnetization are only seen in the gas phase [Fig.~\ref{fig:dynamics}(g)] but they quickly decay in time, whereas in the liquid phase the magnetization immediately relaxes [Figs.~\ref{fig:dynamics}(h, i)].
    
    In summary, the BTC exists only in the region of the phase diagram given by $0\leq\eta\leq1$, $0\leq\chi<1$. The amplitude of oscillation of $m_z$ at infinite times has a finite value in this region due to the persistent BTC oscillations, and are zero everywhere else, with a finite discontinuity at the boundary. So, while the fixed point changes in a continuous way across the boundaries of the BTC region, the amplitude of the BTC oscillations changes discontinuously, featuring a first-order transition.
    
    \begin{figure}[t]
    \centering
    \includegraphics[width=\columnwidth]{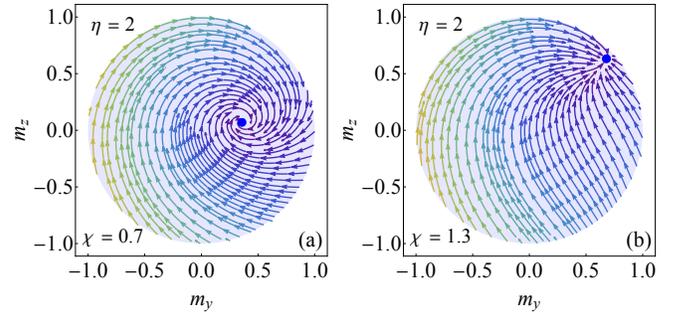}
    \caption{Phase-space flow portraits in the plane $m_x = 0$ for $\eta = 2$. Panel (a): $\chi = 0.7$; panel (b): $\chi = 1.3$. The fixed points are marked in blue. The shaded area corresponds to the region where $\mathcal{N} \le 1$.}
    \label{fig:portraits-eta-2}
\end{figure}
    
Focusing on the gas phase, for $\chi = 0.7$, we see that the magnetization immediately starts to oscillate and the oscillations decay in time [Fig.~\ref{fig:dynamics}(d)]. We checked that the oscillations exponentially decay in time, with the amplitude of the envelope following the law $ A(t) = A_0 \exp[-B(\eta) Jt]$. For $\eta\to1^+$, the decay rate $B(\eta)=0.7 (\eta-1)^2$ as shown\ in Fig.~\ref{fig:metastable}, hinting to long-lived oscillations near the BTC phase.  For $\eta = 1$, one falls in the BTC phase, the oscillations are persistent, and $B(1) = 0$.
    \begin{figure*}
       \centering
       \includegraphics[width=\textwidth]{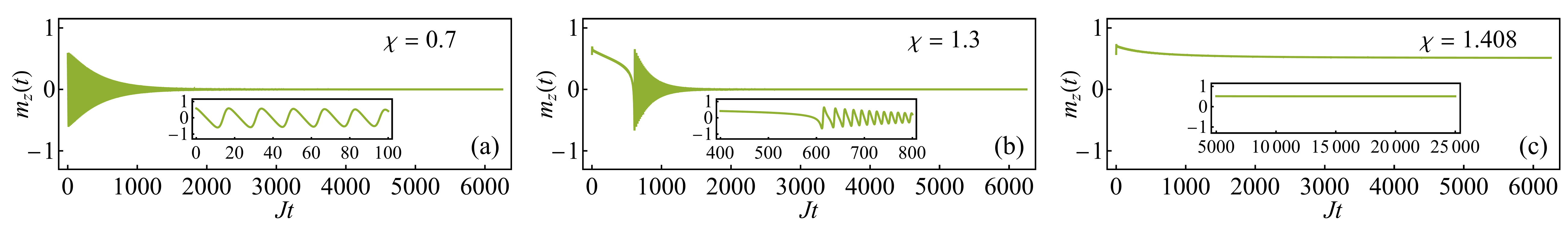}
       \caption{Time evolution of the magnetization for $\eta = 1.1$ for several values of the dissipation strength $\chi$. (a): $\chi = 0.7$; (b): $\chi = 1.3$; (c): $\chi = 1.408$. The insets of panels (a) and (b) zoom in on the regions where oscillations of $m_z$ start to occur. The inset of panel (c) shows the behavior of $m_z$ at longer times: for this choice of $\chi$, the magnetization does not oscillate.}
       \label{fig:phase-diagram}
   \end{figure*}
    
    In order to better understand the short-range regime, let us consider some phase space portraits of the dynamics in the plane $m_x=0$. For small values of $\chi$, the steady-state magnetization is close to zero and the system falls to the fixed point following rapidly decaying oscillations, signaled by spiraling orbits in the phase space portraits in the plane $m_x = 0$ [Fig.~\ref{fig:portraits-eta-2}(a), compare with Fig.~\ref{fig:dynamics}(g)]. This is the gas-like phase shown in the leftmost part of Fig.~\ref{fig:sketch}. Instead, for larger values of $\chi$ we are in the liquid-like phase (see Fig.~\ref{fig:sketch}), and the system converges to a fixed point with large $m_z$ with no oscillations [Fig.~\ref{fig:portraits-eta-2}(b), compare with Fig.~\ref{fig:dynamics}(i)].
    
    \textit{Coexistence region.} Some interesting dynamical behaviors occur when $\eta$ is close to one, inside the coexistence region of the phase diagram. Here, there is a regime where the magnetization starts oscillating around $m_z = 0$ after a delay.  As an example, here we discuss the case of $\eta = 1.1$. For three choices of $\chi$, we numerically solve Eqs.~\eqref{eq:power-law-lindblad-magnetization-mf} and plot the time evolution of $m_z(t)$ in Figs.~\ref{fig:phase-diagram}(a--c).  %In order to better understand this phenomenon, one has to look at the fixed-point magnetization $m_z(\infty)$ versus $\chi$, that we plot in Fig.~\ref{fig:phase-diagram}(a) (curves $\eta=1.01\,\eta=1.1\,\eta=1.5$).

    For $1<\chi\le 1.4$, $m_z(t)$ starts to oscillate at a later time. The time at which oscillations start becomes longer with $\chi$ approaching $1.4$. It also becomes longer for $\eta\to1^+$ since, for $\eta = 1$, the system goes back to the infinite range case where, for $1<\chi\le 1.4$, the system is already in the magnetized phase and no oscillations are visible in $m_z(t)$. This behavior is shown in Figs.~\ref{fig:phase-diagram}(b) for $\chi = 1.3$. Right after the critical point, the oscillations in $m_z(t)$ disappear completely, see Fig.~\ref{fig:phase-diagram}(c) where we show the case of $\chi = 1.408$. In all cases, there is no trace of the persistent oscillations indicating a BTC phase, at least at the mean-field level of the analysis thus far. The endpoint of this region tends to $\chi = \sqrt{2}$ when $\eta\to1^+$.

\begin{figure}
        \centering
        \includegraphics[width=0.6\columnwidth]{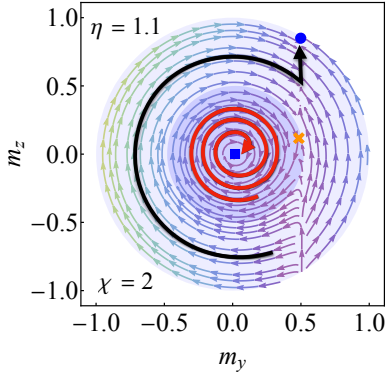}
        \caption{Phase-space flow in the plane $m_x = 0$ for $\eta = 1.1$ and $\chi = 2$. The attractive fixed points are marked in blue with a square ($\mathcal{N} \approx 0$) and a circle ($\mathcal{N} \approx 1$). The orange cross marks the unstable fixed point. The light shaded area corresponds to the region where $\mathcal{N} \le 1$. The dark shaded area denotes the basin of attraction of the fixed point with small $\mathcal{N}$. Two trajectories, where the initial states fall within the basin of attraction of the two stable fixed points, are shown as an example.}
        \label{fig:portraits-eta-1.1}
    \end{figure}

    In order to explain the transient seen in Fig.~\ref{fig:phase-diagram}, we have to  consider that we are in the coexistence regime, where there are two stable fixed points (and an unstable one). Each of the stable fixed points has his own basin of attraction, and the dynamics eventually reaches one or another according to the chosen initial conditions. By studying the phase-space portraits in this regime, we see that, if the initial state has a total magnetization $\mathcal{N}$ [Eq.~\eqref{eq:conserved}] above a certain threshold, the fixed point with $\mathcal{N} \approx 0$ is never reached. We can see this in the phase-space portrait shown in Fig.~\ref{fig:portraits-eta-1.1} for $\eta = 1.1$ and $\chi = 2$ in the plane $m_x = 0$. The system converges to the fixed point $m_z^{(1)}$ (black line) if the initial state falls within the basin of attraction of this point, i.\,e., if the initial total magnetization is larger than $\mathcal{N} = 0.5$. Instead, if $\mathcal{N} < 0.5$, the system spirals towards the fixed point with $\mathcal{N} \approx 0$.
    
     The small-$\mathcal{N}$ fixed point is continuously connected to  the one in Fig.~\ref{fig:portraits-eta-2}(a), and the spiraling dynamics of the trajectories are similar. The large-$\mathcal{N}$ fixed point is continuously connected to the one in Fig.~\ref{fig:portraits-eta-2}(b) and trajectories converge to it without spiraling. The coexistence of fixed points with these properties of the trajectories converging to them explains the results shown in Fig.~\ref{fig:phase-diagram}. Here the system is initialized inside the basin of attraction of the small-$\mathcal{N}$ fixed point, and near the boundary between the two attraction basins. So, there is a transient where the system behaves as it was converging to the large-$\mathcal{N}$ fixed point, and then the trajectory spirals around the small-$\mathcal{N}$ fixed point and converges to it.

\section{Quantum correlations: third-order cumulant expansion}
\label{sec:third-cumulant}
    
\footnotetext[2]{Notice that the case $j=l$ is trivial, due to the Pauli-matrix algebra ${\sigma}_j^\alpha{\sigma}_j^\beta = 2i\epsilon^{\alpha\beta\gamma}{\sigma}_j^\gamma$, with $\epsilon^{\alpha\beta\gamma}$ the Ricci fully antisymmetric tensor.}
    
Eqs.~\eqref{eq:power-law-lindblad-magnetization-mf} are obtained performing the mean-field approximation, which is equivalent to neglect quantum fluctuations $\braket{{\sigma}_j^\alpha{\sigma}_l^\beta}_t\simeq \braket{{\sigma}_j^\alpha}_t\braket{{\sigma}_l^\beta}_t$ $\forall j,l = 1,\,\dots,\,N$ with $j\neq l$, and $\alpha,\beta=x,y,z$~\cite{Note2}. Here we write $\braket{(\cdots)}_t\equiv \Tr[\rho(t)(\cdots)]$ for brevity. This is a strong approximation, because from a physical point of view it is equivalent to state that the density matrix $\rho(t)$ gives rise to distributions of measurement outcomes of ${\sigma}_j^\alpha$ with no fluctuations. %We ask ourselves if this is a good approximation, when one wants to study the expectations of the collective spins in the thermodynamic limit, as we are doing in Eqs.~\eqref{eq:power-law-lindblad-magnetization-mf} [see Eq.~\eqref{mm:eqn}]. 
This is a good approximation in the case of $\eta = 0$ (see Ref.~\cite{iemini-prb-2021:btc-collective-d-level-systems}),\delete{and Appendix~A}as well as in the case $\eta\to\infty$ (see Ref.~\cite{maghrebi-pra-2021:driven-dissipative-ising-model}). By contrast its validity is not known  for intermediate values of $\eta$. \addition{Here, we focus our attention on the long-range regime $0 \le \eta \le 1$.}

%Results on the absence of quantum chaos for any value of $\chi$ and $\eta$ (see App.~\ref{app:lyap}) suggest that also for $1<\eta<\infty$ fluctuations are irrelevant and the mean field is exact. Indeed, if the largest Lyapunov exponent is vanishing and all initial conditions give rise to the same asymptotic state, there is no mechanism to amplify initially small quantum fluctuations, as occurs in contrast in chaotic systems where nearby trajectories diverge exponentially in time. In order to put this observation on a firmer quantitative ground, we can explicitly include quantum fluctuations using a third order cumulant expansion~\cite{hagele-prb-2018:higher-order-moments-cumulants}. 

    \begin{figure}[t]
        \centering
        \includegraphics[width=\columnwidth]{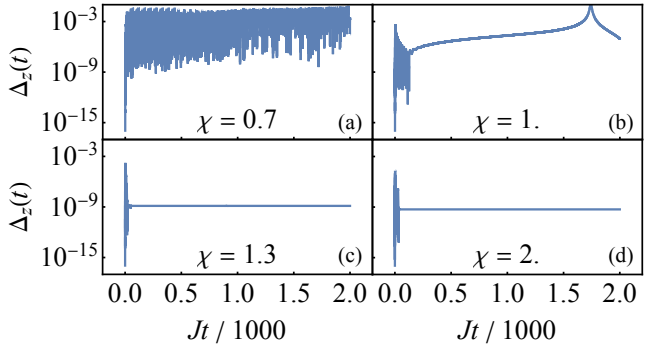}
        \caption{Variance $\Delta_z = C_{zz} - m_z^2$ versus time, for \replace{$\eta = 1.1$}{$\eta = 0.5$} and several values of $\chi = \gamma/4J$, in the Gaussian approximation and the thermodynamic limit. The initial condition is $m_z(0) = C_{zz} = 1$, thus $\Delta_z(0) = 0$. Panel (a): $\chi = 0.7$; Panel (b): $\chi = 1$; Panel (c): $\chi = 1.3$; Panel (d): $\chi = 2$.}
        \label{fig:variance}
    \end{figure}

In order to clarify the relevance of quantum fluctuations,  we apply a third order cumulant expansion~\cite{hagele-prb-2018:higher-order-moments-cumulants}. %This is a further refinement of the mean field approximation we have considered so far.
In the mean field approximation one considers the distribution of the measurement outcomes of ${\sigma}_j^\alpha$ and imposes that the nontrivial second cumulants are vanishing, $C_2(\alpha,\beta,t) \equiv\braket{{\sigma}_j^\alpha{\sigma}_l^\beta}_t-\braket{{\sigma}_j^\alpha}_t\braket{{\sigma}_l^\beta}_t =0$, with $j\neq l$~\cite{Note2}. Notice that $C_2$ is independent of $j$ and $l$, \replace{due to invariance under permutations}{a valid assumption in the thermodynamic limit in the long-range regime, see Appendix~\ref{app:correlations}}. 
The mean-field Eqs.~\eqref{eq:power-law-lindblad-magnetization-mf} can be obtained from the exact equations of motion [see Eqs.~\eqref{eq:magnetization-exact}] requiring that the second cumulants of the probability distributions of the magnetization components vanish.

In the third-order cumulant expansion, instead, one imposes that the nontrivial third cumulants (and all the next ones) are vanishing, that is to say 
\begin{align} \label{C3:eqn}
  C_3(\alpha,\beta,\gamma,t)&=\braket{{\sigma}_j^\alpha{\sigma}_l^\beta{\sigma}_m^\gamma}_t-
  \braket{{\sigma}_j^\alpha{\sigma}_l^\beta}_t\braket{{\sigma}_m^\gamma}_t\notag\\
  &\quad-\braket{{\sigma}_j^\alpha{\sigma}_m^\gamma}_t\braket{{\sigma}_l^\beta}_t
   -
  \braket{{\sigma}_l^\beta{\sigma}_m^\gamma}_t\braket{{\sigma}_j^\alpha}_t\notag\\
  &\quad+
  2\braket{{\sigma}_l^\beta}_t\braket{{\sigma}_m^\gamma}_t\braket{{\sigma}_j^\alpha}_t=0\,.
\end{align}
This is the so-called Gaussian approximation, because one assumes that in the distribution of the measurement outcomes of ${\sigma}_j^\alpha$ only the first two cumulants are nonvanishing, as appropriate for Gaussian distributions.

     Assuming the  Gaussian approximation Eq.~\eqref{C3:eqn}, the equations of motion are more involved and include also the correlations. Thanks to \addition{the effective} permutation invariance \addition{of the model in the thermodynamic limit in the long-range regime---see discussion in Appendix~\ref{app:correlations}, one reduces to solve a ODE system with nine equations} [see Eqs.~\eqref{eq:third-cumulant}]. %This is equivalent to disregarding correlations, since if $C_2(\sigma^\alpha_i, \sigma^\beta_j) = 0$, then  $ \langle \sigma^\alpha_i \sigma^\beta_j \rangle = \langle \sigma^\alpha_i\rangle \langle \sigma^\beta_j\rangle $. This allows writing a closed-form system of three nonlinear ordinary differential equations for the $m_\alpha$'s. 

%    However, the exact equations of motion for the magnetization, Eqs.~\eqref{eq:magnetization-exact}, depend on $2$-body quantum correlations between spin components. This is a general rule, since the dynamical equations of $k$-body spin operators include terms with $(k+1)$-body correlation functions, thus writing down a closed-form system of differential equations for the exact dynamics of the spin system is impossible in the thermodynamic limit.

    %One way to keep some information regarding correlation is by requiring that the third cumulant is zero instead. This is the Gaussian approximation, since Gaussian distributions are entirely specified by the first two cumulants, i.\,e., mean values and covariances. In this way, the $3$-body correlation functions appearing in the dynamical equations of the $2$-body correlation functions can be decomposed as $\langle x_1 x_2 x_3 \rangle = \langle x_1 x_2\rangle \langle x_3\rangle + \langle x_1 x_3\rangle \langle x_2\rangle + \langle x_2 x_3\rangle \langle x_1 \rangle - 2 \langle x_1 \rangle \langle x_2\rangle \langle x_3\rangle$, hence it is possible to write down a closed-form system of ordinary differential equations for the magnetization and $2$-body correlations only. 
    
    \addition{We numerically solve these equations initializing the} system in the uncorrelated state with $m_z(0) = 1$, \addition{with vanishing second cumulants}. Letting the system evolve, it has the freedom to develop nonvanishing second-order cumulants, and then get nontrivial quantum correlations. Taking $0\leq\eta\leq1$, whatever are the considered values of $\chi$, we find that these second-order cumulants \addition{take values of order of $10^{-2}$ or} smaller.\delete{than $10^{-3}$}We show this fact in Fig.~\ref{fig:variance}, where we plot some examples of evolution of $\Delta_z(t) = C_{2}(z,z,t)$. In this figure we focus on the case of \replace{$\eta = 1.1$}{$\eta = 0.5$} as an example of \replace{short}{long} range, but we have checked that the results discussed here are representative of all other values of $\eta$. 
    
    The\addition{se findings imply that} results obtained with the mean field \addition{in the long-range regime $0<\eta\le1$ remain} almost unchanged when quantum fluctuations are considered, because these fluctuations are vanishing small, even if the system is allowed to develop them \addition{at the level of second cumulants}. %This implies that the mean field is an excellent approximation in the thermodynamic limit, and the results discussed in Sec.~\ref{sec:mean-field} are essentially exact. 
    We emphasize that this result is valid in the thermodynamic limit, where the terms that break permutation invariance in the dynamical equations for the correlations disappear when $0<\eta<1$ (see Appendix~\ref{app:correlations}). Further studies are needed to better understand it from the analytical point of view. 
    
\begin{figure}[t]
        \centering
        \includegraphics[width=0.95\columnwidth]{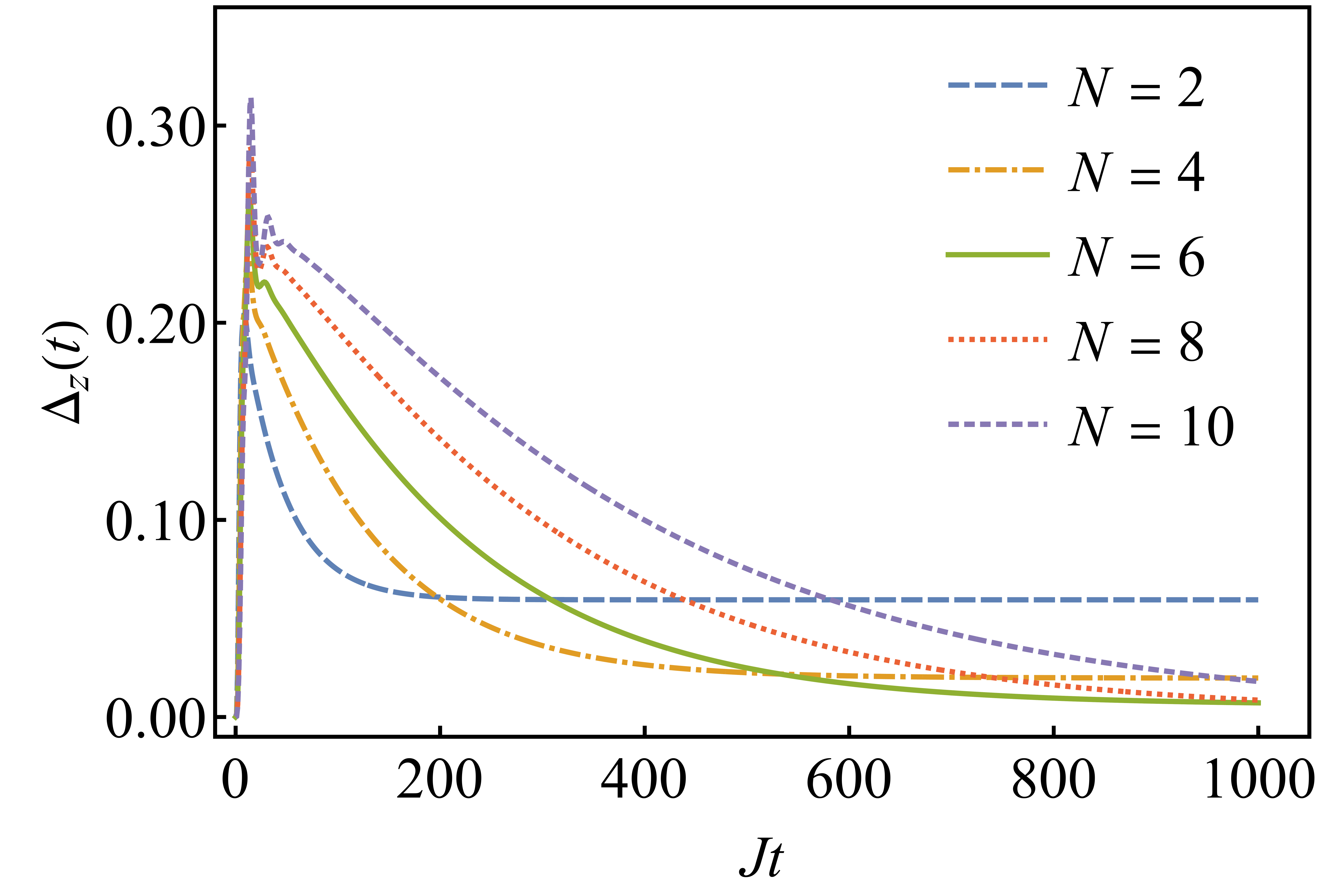}
        \caption{Variance $\Delta_z = C_{zz} - m_z^2$ versus time, for \replace{$\eta = 1.1$}{$\eta = 0.5$} and $\chi = 0.7$, for small system sizes. The initial condition is the factorized density matrix with each spin in the eigenstate of the corresponding $\hat{\sigma}_j^z$ with positive eigenvalue. \maybedelete{In the inset we show the scaling of the maximum of $\Delta_z(t)$ as a function of $N$.}}
        \label{fig:variance-finite-n}
    \end{figure}

    The second-order cumulants do not vanish in finite-size systems. In this case we see that finite second-order cumulants (and the associated quantum fluctuations) develop in time. We show this in Fig.~\ref{fig:variance-finite-n} for the case of \replace{$\eta = 1.1$}{$\eta = 0.5$} and $\chi = 0.7$, where we plot $\Delta_z(t)$ versus $t$, obtained by numerically solving the full Lindblad equation for finite $N$. \maybedelete{In the inset we use the Gaussian approximation at finite system size $N$ [see Eq.~\eqref{eq:third-cumulant-finite-size}], where we keep terms proportional to $1/N$ to take into account finite-size effects, and evaluate the maximum over the dynamics of $\Delta_z(t)$. Plotting this quantity $\max\Delta_z$ versus $N$ we see that it converges to the $N\to\infty$ value for large system sizes, but at small sizes it is order 1. So, at finite size, the Gaussian approximation is bad and further orders in the cluster expansion are needed. In the main panel of Fig.~\ref{fig:variance-finite-n} we plot $\Delta_z(t)$ versus $t$ obtained by numerically integrating the Lindblad Eq.~\eqref{eq:lindbladian-interpolated} for small $N$. We see that also in absence of any approximation $\Delta_z(t)$ reaches values of order 1 at finite size.}
    
    %$C_{\alpha\beta}(0) = \delta_{\alpha,z}\delta_{\beta,z}$. This choice enforces a vanishing second cumulant, thus the dynamical equations at the mean-field level would ensure zero variance $\Delta_z(t) = C_{zz}(t) - m_z^2(t)$ throughout the evolution. By contrast, at the third cumulant order, $\Delta_z(t)$ can become nonzero, but we expect it to remain small in the regimes where correlations only provide a small correction to the mean-field dynamics. 
    
%    Our results are summarized in Fig.~\ref{fig:variance}. We see that the variance remains small throughout the evolution for all choices of parameters in the intermediate-range regime. On the one hand, this confirms that the mean-field analysis carried out in this paper is valid in all regimes in the thermodynamic limit. On the other hand, since correlations do not significantly alter the mean-field dynamics, the existence of a boundary time crystal phase in the intermediate-range dissipative regime is unlikely.
%----------------------------------------------------------------------------------------------------------------------------------------%
\section{Conclusions} 
\label{sec:conclusions}

%In conclusion, we have considered a boundary time crystal, and we have inquired if the persistent collective oscillations in the thermodynamic limit giving rise to the time-translation symmetry breaking are robust if the strong rotational symmetry is removed. This symmetry amounts to the conservation of the square of the collective spin operator. In order to remove this symmetry, we have considered a simple model of boundary time crystal of Ref.~\cite{iemini-prl-2018:boundary-time-crystals}, and we have replaced the collective spin operators in the Lindbladian with spatially decaying operators. These operators decay as a power law with exponent $\eta$. The class of model dissipation that we analyze can be realized with Rydberg atoms in cavities, as recently discussed in Refs.~\cite{seetharam_2022a,seetharam_2022b}. 

In conclusion, we have introduced a new class of long-range dissipative models that support a time-crystal phase, where the Lindblad operators decay as a power law with exponent $\eta$. The steady-state phase diagram of the model we considered appears quite rich with different transition lines and a coexistence phase. The form of long-range dissipation has been motivated by a recent proposal~\cite{seetharam_2022a,seetharam_2022b} and can be implemented with Rydberg atoms in cavities.

In this model, as opposed to the one proposed in Ref.~\cite{iemini-prl-2018:boundary-time-crystals}, the square of the collective spin operator is no longer conserved. Nevertheless, studying the dynamics of the total magnetization, we see that the time-translation symmetry breaking oscillations persist if $0<\eta\leq 1$ (long-range regime). In this regime there is a transition between a small-$\chi$ time-translation symmetry breaking phase and a large-$\chi$ phase with no time crystal, where the magnetization attains an asymptotic finite value. For $0<\eta\leq 1$ the transition point in $\chi$ is independent of $\eta$. So, in this regime, the thermodynamic-limit dynamics is independent of $\eta$, similarly to what happens in Hamiltonian models with long-range interaction. Remarkably, although the square of the collective spin operator is not conserved for any finite size, its expectation in the thermodynamic limit (the square of the total magnetization) is conserved.

Outside this range of parameters, the system reaches an asymptotic steady state. Here, an interesting regime occurs for $1<\eta\leq 1.625$, where the system shows a first-order phase transition line (a discontinuity) of the $z$-magnetization in the $\chi$-$\eta$ plane and there is a region where two stable steady states coexist. This phase transition line terminates with a critical point, beyond which the magnetization is analytic. The way the first-order phase transition develops corresponds to an asymmetric pitchfork bifurcation.

In order to study the thermodynamic-limit dynamics of the total magnetization, we use the mean-field approximation. That means that we impose zero quantum fluctuations  in the Ehrenfest equations of the expectation of the magnetization. From a mathematical point of view, this is equivalent to impose that the distribution of the outcomes of the measurements of the spin components on the wave function has vanishing second cumulant (that is to say no fluctuations). It was already known that this approximation is exact in the thermodynamic limit for $\eta=0$~\cite{iemini-prl-2018:boundary-time-crystals} and $\eta\to\infty$~\cite{maghrebi-pra-2021:driven-dissipative-ising-model}. Here we numerically prove that it is \replace{exact}{quantitatively accurate}, \replace{up to the third cumulant}{in the thermodynamic limit and in the Gaussian approximation}, also for \replace{$1<\eta<\infty$}{$0<\eta\le1$}.

%For that purpose, we first \textcolor{blue}{that there is no chaos, because the system can only attain asymptotic steady states or limit cycles and there are no strange attractors. The absence of chaos, confirmed by the vanishing of the largest Lyapunov exponent, implies that} there is no exponential divergence in time of nearby trajectories. As a consequence, small initial quantum fluctuations are not amplified by the dynamics, and this suggests that the mean-field equations are exact for this specific model. 

In order to do that, we study the Ehrenfest equations beyond the mean-field approximation: imposing that the third-order cumulants are vanishing. So, we approximate that the distribution of the outcomes of the measurements of the spin components on the wave function is a Gaussian. In this approximation, the system has the freedom to build up a nonvanishing second-order cumulant (with the associated fluctuations and correlations). \addition{We focus on the thermodynamic limit for $0\leq \eta \leq 1$, and numerically observe that no second-order cumulant appears, so the  mean-field picture is exact in this regime of parameters. Beyond this range of $\eta$, the second-order cumulant dynamics becomes much more complicated to analyze, because one loses the effective permutation symmetry that emerges in the thermodynamic limit for $0\leq \eta \leq 1$. Studying the correlations in this regime will be the subject of a future publication.} %When we consider finite-size cases, the dynamics can build up nonvanishing second-order cumulants.

Perspectives of future work include also to better understand analytically this absence of second-order cumulants, that we verify here only numerically. Moreover, it will be interesting to break the full permutation symmetry of the model (for instance applying disorder to the Lindbladians), in order to see if the time-crystal phase is stable to this further reduction of symmetry. In this context, another direction to pursue is using an interacting Hamiltonian, in order to see the effect on the BTC of the correlations induced by the interaction. This can be numerically studied by applying the discrete truncated Wigner approximation~\cite{PhysRevX.5.011022,Mink_22,PhysRevLett.128.200602}. Finally, we plan to apply the long-range dissipation to a model displaying quantum chaos with an infinite-range dissipator~\cite{Hartmann_2017}. Our aim is to understand how the chaotic properties change, and if the mean-field approximation is still valid in the thermodynamic limit when there is chaos.
 %   In conclusion, we studied a dissipative spin model under the action of a Lindbladian that interpolates between infinite-range dissipation and local dissipation. We discovered the emergence of a mean-field boundary time crystal phase in the long-range regime. This is interesting \textit{per se} because this phase was never before observed in a system where Lindblad operators are not uniform superpositions of single spin operators. This model does not support any other time crystal phase for shorter-range dissipation, neither at the mean-field level nor when correlation effects are considered at the third order in the cumulant expansion.

\begin{acknowledgments}
This work has been funded by project code PIR01\_00011 “IBiSCo”, PON 2014-2020. The work has been supported by the ERC under grant  agreement n.~101053159 (RAVE). We thank \addition{F.~Iemini and} M.~M. Wauters for useful comments on the manuscript.
\end{acknowledgments}

\appendix
\section{Kac normalization}
\label{app:kac}

    The power-law Lindblad operators are defined in Eq.~\eqref{eq:power-law-lindblad}, where
    \begin{equation}
        f_{ij}(\eta) = \frac{K^{(N)}(\eta)}{{[\min(\lvert i - j \rvert, N - \lvert i - j \rvert) + 1]}^\eta}.
    \end{equation}
    The \addition{Kac} normalization factor $K^{(N)}(\eta)$ is defined in such a way that $\sum_{j=1}^N f_{ij}(\eta) = 1$ \replace{This relation must hold for all $i$ due to the system being permutation invariant}{for all $i$}. For convenience, let us then suppose $N$ is even and fix $i = N/2$, so that the denominator of $f_{ij}(\eta)$ contains the term $N/2-j$ if $j\le N/2$, and $j - N/2$ if $j > N/2$. Thus, we have
    \begin{align}
        \sum_{j=1}^N f_{ij}(\eta) &= K^{(N)}(\eta) \Biggl[\,\sum_{j=1}^{N/2} \frac{1}{{(N/2-j+1)}^\eta} \notag\\
        &\qquad+ \sum_{j=N/2+1}^{N} \frac{1}{{(j-N/2+1)}^\eta}\,\Biggr] \notag \\
        &= K^{(N)}(\eta) \left[\,\sum_{k=1}^{N/2} \frac{1}{k^\eta} + \sum_{k=2}^{N/2 + 1} \frac{1}{k^\eta}\,\right] \notag \\
        &= K^{(N)}(\eta) \left[\,2\sum_{k=1}^{N/2} \frac{1}{k^\eta} - 1 + {\left(\frac{2}{N+2}\right)}^\eta\,\right] = 1.
    \end{align}
    Similar algebra holds for the case of odd $N$. For even $N$, we find 
    \begin{equation}
        K^{\text{(even $N$)}}(\eta) = \frac{1}{2 H_{1+N/2}^{(\eta)} - 1 - {{\left(1+\frac{N}{2}\right)}^{-\eta}}},
    \end{equation}
    where $H_n^{(k)} = \sum_{j=1}^n j^{-k}$ is the $n$-th Harmonic number of order $k$. Similarly, for odd $N$:
    \begin{equation}
        K^{\text{(odd $N$)}}(\eta) = \frac{1}{2 H_{1+(N-1)/2}^{(\eta)} - 1}.
    \end{equation}
%----------------------------------------------------------------------------------------------------------------------------------------%
\section{Mean-field magnetization dynamics}
\label{app:magnetization-dynamics}

    When the time evolution is described by the Lindbladian of Eq.~\eqref{eq:lindbladian-interpolated}, then for any observable $O$ we can write the dissipative Ehrenfest theorem as
\begin{equation}
    \dot{\langle O \rangle} = i \langle [H, O] \rangle + \frac{\gamma}{2} \sum_{i=1}^{N} \langle L_i^\dagger [O, L_i] + [L_i^\dagger, O] L_i\rangle,
\end{equation}
where we avoid to write the $\eta$- and time-dependence for convenience and the expectation values are taken over $\rho(t)$. Evaluating this equation for the magnetization components $m_\alpha = 2 \langle S_\alpha\rangle  / N$, we obtain the following equations,
\begin{align}
    \dot{m}_x &= -\frac{\gamma}{2N} \sum_{i=1}^N \langle L_i^\dagger M_i + M_i L_i \rangle,\\ \dot{m}_y &= U_y^{(N)} + \frac{\gamma}{i \, 2N} \sum_{i=1}^N \langle L_i^\dagger M_i - M_i L_i \rangle,\\
    \dot{m}_z &= U_z^{(N)} - \frac{2\gamma}{N} \sum_{i=1}^N \langle L_i^\dagger L_i\rangle,
\end{align}
where we defined $M_i(\eta) = \sum_{ij} f_{ij}(\eta) \sigma^z_i$ and $U_\alpha^{(N)} = 2i\, \langle [H, S_\alpha] \rangle / N$. In terms of single-spin operators, these equations can be rewritten as follows:
\begin{gather}
    \dot{m}_x = -\frac{\gamma}{2N} \sum_{j=1}^N \mathcal{F}_{jj}(\eta) \, m_x - \frac{\gamma}{2N} \sum_{j=1}^N \sum_{k\ne j}^N \mathcal{F}_{jk}(\eta) \, C_{jk}^{xz},\notag\\ 
    \dot{m}_y = U_y^{(N)} -\frac{\gamma}{2N} \sum_{j=1}^N \mathcal{F}_{jj}(\eta) \, m_y - \frac{\gamma}{2N} \sum_{j=1}^N \sum_{k\ne j}^N \mathcal{F}_{jk}(\eta) \, C_{jk}^{yz},\notag\\
    \begin{aligned}
    \dot{m}_z &= U_z^{(N)} + \frac{\gamma}{N} \sum_{j=1}^N \mathcal{F}_{jj}(\eta) \, (1-m_z) \notag\\ &\quad+\frac{\gamma}{2N} \sum_{j=1}^N \sum_{k\ne j}^N \mathcal{F}_{jk}(\eta) \, (C_{jk}^{xx} + C_{jk}^{yy}),
    \end{aligned}
\end{gather}
where we introduced the correlation functions $C_{jk}^{\alpha\beta} = \langle \sigma_j^\alpha \sigma_k^\beta\rangle$ and the coefficients $\mathcal{F}_{jk}(\eta) = \sum_{i=1}^N f_{ij}(\eta) \, f_{ik}(\eta) $. They satisfy $\sum_{j=1}^N \sum_{k=1}^N \mathcal{F}_{jk} = N$. Due to the form of $f_{ij}$, the coefficients $\mathcal{F}_{jk} $ only depend on $\lvert j-k\rvert$, in particular $F_\eta^{(N)} \equiv \mathcal{F}_{jj} $ does not depend on the spacial index. Thus, the equations of motion can be simplified to read
\begin{gather}\label{eq:magnetization-exact}
    \dot{m}_x = -\frac{\gamma}{2} F_\eta^{(N)} \, m_x - \frac{\gamma}{2N} \sum_{j=1}^N \sum_{k\ne j}^N \mathcal{F}_{jk}(\eta) \, C_{jk}^{xz},\notag\\ 
    \dot{m}_y = U_y^{(N)} -\frac{\gamma}{2} F_\eta^{(N)} \, m_y - \frac{\gamma}{2N} \sum_{j=1}^N \sum_{k\ne j}^N \mathcal{F}_{jk}(\eta) \, C_{jk}^{yz},\\
    \begin{aligned}
    \dot{m}_z &= U_z^{(N)} + \gamma \, F_\eta^{(N)} \, (1-m_z) \notag \\&\quad {} + \frac{\gamma}{2N} \sum_{j=1}^N \sum_{k\ne j}^N \mathcal{F}_{jk}(\eta) \, (C_{jk}^{xx} + C_{jk}^{yy}).
    \end{aligned}
\end{gather}

In the mean field approximation, we suppose that the system state is uncorrelated and can be written as a tensor product of single-qubit density matrices (Gutzwiller ansatz): $\rho = \bigotimes_{i=1}^N \rho_i$. Thus, we assume $C^{\alpha\beta}_{jk} = m_\alpha m_\beta $ due to translational invariance and obtain the mean-field equations of motion in the thermodynamic limit
\begin{gather}
    \dot{m}_x = -\frac{\gamma}{2} F_\eta \, m_x - \frac{\gamma}{2} (1-F_\eta) \, m_{x} m_{z},\notag\\ 
    \dot{m}_y = U_y^{(\infty)} -\frac{\gamma}{2} F_\eta \, m_y - \frac{\gamma}{2} (1-F_\eta) \, m_y m_{z},\notag\\ 
    \dot{m}_z = U_z^{(\infty)} + \gamma \, F_\eta \, (1-m_z) + \frac{\gamma}{2} (1-F_\eta) \, (m_{x}^2 + m_{y}^2),
\end{gather}
also shown in the main text [Eqs.~\eqref{eq:power-law-lindblad-magnetization-mf}].

To derive these equations in the thermodynamic limit $N\to\infty$, we analytically evaluated the coefficients $F_\eta$. In particular, we have that
\begin{align}\label{eq:coeff-f1}
    F_\eta &= \lim_{N\to\infty} F^{(N)}_\eta \notag\\
    &=\lim_{N\to\infty} \sum_{i=1}^N \frac{{[K^{(N)}(\eta)]}^2}{{(\min{\lvert i-j\rvert, N-\lvert i - j \rvert} + 1)}^{2\eta}}.
\end{align}
This relation must hold for all $j$, thus we can fix $j = N/2$ and follow the same steps as those of Appendix~\ref{app:kac}. In this way, we easily obtain
\begin{equation}
    F_\eta =\begin{cases}
        \displaystyle\frac{2\zeta(2\eta)-1}{{[2\zeta(\eta) - 1]}^2} & \text{$\eta > 1$,}\\
			0	& \text{$0\le\eta\le1$,}
    \end{cases}
\end{equation}
where $\zeta(z) = \lim_{n\to\infty} H_n^{(z)}$ is Riemann's Zeta function, defined for $\operatorname{Re}(z) > 1$. For finite $N$, the sum in Eq.~\eqref{eq:coeff-f1} can be evaluated numerically.

\begin{widetext}
\section{Correlations dynamics in the Gaussian approximation}\label{app:correlations}

As mentioned in the main text, one way to keep some information regarding correlation is by requiring that the third cumulant is zero. This is known as the Gaussian approximation. The $3$-body correlation functions appearing in the dynamical equations of the $2$-body correlation functions can be decomposed as $\langle x_1 x_2 x_3 \rangle = \langle x_1 x_2\rangle \langle x_3\rangle + \langle x_1 x_3\rangle \langle x_2\rangle + \langle x_2 x_3\rangle \langle x_1 \rangle - 2 \langle x_1 \rangle \langle x_2\rangle \langle x_3\rangle$, hence it is possible to write down a closed-form system of ordinary differential equations for the magnetization and $2$-body correlations only. In the case of permutation invariant systems, this ODE system has nine equations.

When the system Hamiltonian is $ H = 2J S_x$, the time evolution of the correlation function $C_{lm}^{\alpha\beta} = \langle \sigma_l^\alpha \sigma_m^\beta\rangle$ in the power-law dissipative model in the Gaussian approximation is given by
%\begin{widetext}
\begin{align}\label{eq:third-cumulant-finite-size}
    \dot{C}_{lm}^{\alpha\beta} &= 2J (\epsilon_{x \alpha \theta} C^{\theta \beta}_{lm} + \epsilon_{x \beta \theta} C^{\alpha \theta}_{lm})\notag \\
    &{}+ \frac{\gamma}{2}\biggl\lbrace \epsilon_{\beta x \theta} \Bigl[m_y (C_{lm}^{\alpha \theta} - 2 m_\alpha m_\theta) + m_{\theta} \boldsymbol{\bigl(} \mathcal{F}_{lm}(\delta_{y\alpha} - \epsilon_{x\alpha\zeta} m_{\zeta}) + \sum_{j\ne l} \mathcal{F}_{jm} C_{lj}^{\alpha y} \boldsymbol{\bigr)} + m_\alpha \boldsymbol{\bigl(} \mathcal{F}_{mm}(\delta_{y\theta} - \epsilon_{x\theta\zeta} m_{\zeta}) + \sum_{j\ne m} \mathcal{F}_{jm} C_{jm}^{y\theta} \boldsymbol{\bigr)}\Bigr]\notag\\
    &{}-\epsilon_{\beta y \theta} \Bigl[m_x (C_{lm}^{\alpha \theta} - 2 m_\alpha m_\theta) + m_{\theta} \boldsymbol{\bigl(} \mathcal{F}_{lm}(\delta_{x\alpha} + \epsilon_{y\alpha\zeta} m_{\zeta}) + \sum_{j\ne l} \mathcal{F}_{jm} C_{lj}^{\alpha x} \boldsymbol{\bigr)} + m_\alpha \boldsymbol{\bigl(} \mathcal{F}_{mm}(\delta_{x\theta} + \epsilon_{y\theta\zeta} m_{\zeta}) + \sum_{j\ne m} \mathcal{F}_{jm} C_{jm}^{x\theta} \boldsymbol{\bigr)}\Bigr]\notag\\
    &{}+\epsilon_{\alpha x \theta} \Bigl[m_y (C_{lm}^{\theta\beta} - 2 m_\beta m_\theta) + m_{\theta} \boldsymbol{\bigl(} \mathcal{F}_{lm}(\delta_{y\beta} - \epsilon_{x\beta\zeta} m_{\zeta}) + \sum_{j\ne m} \mathcal{F}_{lj} C_{jm}^{y\beta} \boldsymbol{\bigr)} + m_\beta \boldsymbol{\bigl(} \mathcal{F}_{ll}(\delta_{y\theta} - \epsilon_{x\theta\zeta} m_{\zeta}) + \sum_{j\ne l} \mathcal{F}_{lj} C_{lj}^{\theta y} \boldsymbol{\bigr)}\Bigr]\notag\\
    &{}-\epsilon_{\alpha y \theta} \Bigl[m_x (C_{lm}^{\theta\beta} - 2 m_\beta m_\theta) + m_{\theta} \boldsymbol{\bigl(} \mathcal{F}_{lm}(\delta_{x\beta} + \epsilon_{y\beta\zeta} m_{\zeta}) + \sum_{j\ne m} \mathcal{F}_{lj} C_{jm}^{x \beta} \boldsymbol{\bigr)} + m_\beta \boldsymbol{\bigl(} \mathcal{F}_{ll}(\delta_{x\theta} + \epsilon_{y\theta\zeta} m_{\zeta}) + \sum_{j\ne l} \mathcal{F}_{lj} C_{lj}^{\theta x} \boldsymbol{\bigr)}\Bigr]\biggr\rbrace.
\end{align}
%\end{widetext}

\addition{Due to the translational invariance of the model, we have $C_{lm}^{\alpha\beta} = C_{\lvert l-m \rvert}^{\alpha\beta}$. In the long-range regime in the thermodynamic limit ($0<\eta\le 1$), due to the Kac factors (see Appendix~\ref{app:kac}), the terms in Eq.~\eqref{eq:third-cumulant-finite-size} containing $\mathcal{F}_{lm}$ and $\mathcal{F}_{mm}$ vanish. As a consequence, in this limit, the equations Eq.~\eqref{eq:third-cumulant-finite-size} become symmetric under all the site-permutation transformations, as it is easy to check. So, initializing the system in a permutation-invariant state---as we do, the correlations become independent of the lattice indices, and the system behaves as it is symmetric under site permutations, resembling the infinite-range case $\eta = 0$.}

\subsection*{Thermodynamic limit}

If we exploit the model's permutation invariance ($C_{lm}^{\alpha\beta} = C_{\alpha\beta}$) \addition{emerging in the long-range regime} and sum both sides of the equation over $l$ and $m\ne l$ (so that the spin operators in the expectation value always commute), we get the following equation in the thermodynamic limit:
%\begin{widetext}
\begin{align}
    \dot{C}_{\alpha\beta} &= 2J(\epsilon_{x \alpha \theta} C_{\theta \beta} + \epsilon_{x \beta \theta} C_{\alpha \theta}) \notag \\
    &{}+ \frac{\gamma}{2}\biggl\lbrace \epsilon_{\beta x \theta} \Bigl[m_y (C_{\alpha\theta}-2m_\alpha m_\theta) + m_\theta C_{\alpha y} + m_\alpha \boldsymbol{\bigl(} F_\eta (\delta_{y\theta} - \epsilon_{x\theta\zeta} m_\zeta) + C_{y\theta}(1-F_\eta)\boldsymbol{\bigr)}\Bigr]\notag\\
    &{}-\epsilon_{\beta y \theta} \Bigl[m_x (C_{\alpha\theta}-2m_\alpha m_\theta) + m_\theta C_{\alpha x} + m_\alpha \boldsymbol{\bigl(} F_\eta (\delta_{x\theta} + \epsilon_{y\theta\zeta} m_\zeta) + C_{x\theta}(1-F_\eta)\boldsymbol{\bigr)}\Bigr]\notag\\
    &{}+\epsilon_{\alpha x \theta} \Bigl[m_y (C_{\theta\beta}-2m_\beta m_\theta) + m_\theta C_{ y\beta} + m_\beta \boldsymbol{\bigl(} F_\eta (\delta_{y\theta} - \epsilon_{x\theta\zeta} m_\zeta) + C_{\theta y}(1-F_\eta)\boldsymbol{\bigr)}\Bigr]\notag\\
    {}&-\epsilon_{\alpha y \theta} \Bigl[m_x (C_{\theta\beta}-2m_\beta m_\theta) + m_\theta C_{x \beta} + m_\beta \boldsymbol{\bigl(} F_\eta (\delta_{x\theta} + \epsilon_{y\theta\zeta} m_\zeta) + C_{\theta x}(1-F_\eta)\boldsymbol{\bigr)}\Bigr]\biggr\rbrace.
\end{align}
%\end{widetext}

Putting everything together, we obtain the following system of ordinary differential equations:
%\begin{widetext}
\begin{gather}
    \dot{m}_x = -\frac{\gamma}{2} F_\eta \, m_x - \frac{\gamma}{2} (1-F_\eta) \, C_{xz},\notag\\ 
    \dot{m}_y = 2J m_z -\frac{\gamma}{2} F_\eta \, m_y - \frac{\gamma}{2} (1-F_\eta) \, C_{yz},\notag\\ 
    \dot{m}_z = -2J m_y + \gamma \, F_\eta \, (1-m_z) + \frac{\gamma}{2} (1-F_\eta) \, (C_{xx} + C_{yy}),\notag\\
    \dot{C}_{xx} = -\gamma \left\lbrace m_x \left[ F_\eta m_x + (2-F_\eta)C_{xz}-2 m_x m_z\right] + m_z C_{xx} \right\rbrace,\notag\\
    \dot{C}_{xy} = 2 J C_{xz} - \frac{\gamma}{2} \left\lbrace m_x \left[F_\eta m_y + (2-F_\eta)C_{yz} - 2m_y m_z\right] + m_y \left[F_\eta m_x + (2-F_\eta)C_{xz} - 2 m_x m_z\right] + 2 m_z C_{xy} \right\rbrace,\notag\\
    \begin{aligned}
        \dot{C}_{xz} &= -2J C_{xy} + \frac{\gamma}{2} \bigl\lbrace m_x \left[2F_\eta(1-m_z) + (C_{xx} + C_{yy}) (1-F_\eta) + 2C_{xx}-2m_x^2+2m_z^2-C_{zz}\right] \\
        &\quad {}+m_y (2 C_{xy} - 2 m_x m_y) - m_z \left[ F_\eta m_x + (2-F_\eta)C_{xz}\right]\bigr\rbrace,
    \end{aligned}\notag\\[1ex]
    \dot{C}_{yy} = 4 J C_{yz} - \gamma \left\lbrace m_y \left[F_\eta m_y + (2 - F_\eta) C_{yz} - 2 m_y m_z\right] + m_z C_{yy}\right\rbrace,\notag\\
    \begin{aligned}
        \dot{C}_{yz} &= 2J (C_{zz} - C_{yy}) + \frac{\gamma}{2} \bigl\lbrace m_x(2C_{xy} - 2 m_x m_y) + m_y \bigl[2 C_{yy} - 2 m_y^2 + 2 F_\eta (1-m_z) \\&\quad {}+ (C_{xx} + C_{yy})(1-F_\eta) - C_{zz} + 2m_z^2\bigr] -m_z \left[F_\eta m_y + (2-F_\eta) C_{yz}\right]\bigr\rbrace,
    \end{aligned}\notag\\[1ex]
    \dot{C}_{zz} = -4J C_{yz} + \gamma \left\lbrace 2 m_x (C_{xz} - m_x m_z) + 2 m_y (C_{yz} - m_y m_z) + m_z \left[2F_\eta(1-m_z)+(C_{xx} + C_{yy})(1-F_\eta)\right] \right\rbrace.\label{eq:third-cumulant}
\end{gather}
%\end{widetext}
It is a stiff nonlinear system, thus its numerical solution is only stable when the evolution time is relatively short~\cite{hindmarsh2005sundials}.
\end{widetext}

%    \bibliography{refs2}
\input{main4.bbl}

\end{document}